\documentclass[10pt]{article}
\usepackage{graphicx}
\usepackage{epsfig}
\usepackage{amssymb}
\usepackage{amsmath}
\usepackage{amsthm}
\newtheorem{lemma}{Lemma}[section]
\newtheorem{theorem}{Theorem}[section]

{\end{description}}
{\end{description}}

{\hfill $\bullet$ \\}

\newcommand{\be}{\begin{equation}}
\newcommand{\ee}{\end{equation}}

\begin{document}
    \title{A Fast Third-Step Second-Order Explicit Numerical Approach To Investigating and Forecasting The Dynamic of Corruption And Poverty In Cameroon}
    \author{Eric Ngondiep}
   \date{$^{\text{\,1\,}}$\small{Department of Mathematics and Statistics, College of Science, Imam Mohammad Ibn Saud\\ Islamic University
        (IMSIU), $90950$ Riyadh $11632,$ Saudi Arabia.}\\
     \text{\,}\\
       $^{\text{\,2\,}}$\small{Hydrological Research Centre, Institute for Geological and Mining Research, 4110 Yaounde-Cameroon.}\\
     \text{\,}\\
        \textbf{Email addresses:} ericngondiep@gmail.com/engondiep@imamu.edu.sa}

    \maketitle
   \textbf{Abstract.}
   This paper constructs a third-step second-order numerical approach for solving a mathematical model on the dynamic of corruption and poverty. The stability and error estimates of the proposed technique are analyzed using the $L^{2}$-norm. The developed algorithm is at less zero-stable and second-order accurate. Furthermore, the new method is explicit, fast and more efficient than a large class of numerical schemes applied to nonlinear systems of ordinary differential equations and can serve as a robust tool for integrating general systems of initial-value problems. Some numerical examples confirm the theory and also consider the corruption and poverty in Cameroon.
   \text{\,} \\
    \text{\,}
   \\
   {\bf AMS Subject Classification (MSC). 65M12, 65M06}.

      \section{Introduction and motivation}\label{sec1}
      Poverty and corruption problems remain the core development challenge on a global level. Poverty is defined as the failure to attain a minimum level of well-being and it is the main issue at which people are confronted. In the literature \cite{10ca,16ca,19ca}, many authors have demonstrated that poverty and inequality have still been the forefront of all economic problems. The required consumption per capita condition to ensure a minimum life standardize is called poverty line and it is computed by the Cost of Basic Needs (CBN) technique. The CBN approach is used by a large class of researchers to estimate the poverty line in many countries \cite{4ca}. Corruption is directly related to poverty and it can be observed as one of the major factor of poverty and a barrier to successfully eradicate it.\\

      Transparency International (TI) has defined corruption as the misuse of entrusted power for private gain whereas the Worth Bank (WB) considered the corruption as the abuse of public office for private benefit. In several nations, the phenomenon of corruption has negative impacts on economy and society. Specifically, it is a powerful instrument to destroy a society, can be considered as a cancer to political development, economic and social activities. Furthermore, corruption greatly improves income inequalities by sharing power imbalances in poor countries. In almost all the sectors, mainly the public sector, corruption deepens poverty by delaying and diverting economic growth and productive programs such as: health care and education at the expense of a wide class of capital intensive projects that should suggest better opportunities to overcome illegal expenditures. The corruption deals in different levels: from the high levels corruption which concerns the highest level of national government to the pretty one, that is, the exchange of very small amounts of money or the granting of minors favors by those in lower positions. Thus, the society wants to uphold the value of a Man for what he is and not what he has. Though there exists a strong relationship between corruption and poverty, it has been observed that policy recommendation to combating corruption and poverty has failed. However, Researchers have demonstrated that fighting corruption is a good remedy to reduce poverty \cite{3nrn,7nrn,8nrn}.\\

      In Cameroon, poverty is much more than just having enough money to meet basic needs including: food, clothing and shelter. More precisely, poverty includes: hunger, lack of shelter, being sick and unable to see a physician, unemployment, fear for the future and living one day at time. In addition, in some rural areas children cannot go to school because of worst life conditions and thus are incapable to read and write. Since 1985, the country has developed a new poverty line of 36000 FCFA (equivalent of US $\$65,5$) per month which does not allow people to access minimum consumption conditions for a standard sized household for basic needs mentioned above. Thus, a large class of individuals live in extreme poverty. For more details about extreme poverty and corruption in Cameroon, we refer the readers to \cite{gnm}. For some years now, a majority of cameroonians have shown very little interest in studying and serving in the country. This great lack of motivation is partly due to unemployment, misery and very low salaries which since 1993 are well half below the poverty line: case of private teachers of primary and secondary schools, caretaker, etc...\\

      From $1978$ to $1983$, the poverty line developed in the country was indicating that less than $1\%$ of the population was considered poor and after this period, the government ignored this restriction and the poverty rate grew. In many Cameroon hospitals such as: Yaound\'{e} central hospital, Douala laquintine hospital, Biyem-Assi district hospital, etc..., patients and newborn babies dead almost every day because no nurse or physician can take care of them. They are extremely poor and not are capable to pay the investigation fees. Furthermore, since $2016$ some people in renting in Yaound\'{e} moved house for a new area, so called Elig-Essomballa, where they lived without electricity during more than a year. After submitting many requests to the department in charge of energy (ENEO) without getting a solution, individuals with modest incomes representing $30\%$ of the population in the district, considered this problem as their own project whereas those living in high level of misery were still in obscurity because unable to contribute to the project even as small as possible. From $2019$ to date, although nothing is done regarding electricity in the area, the energy department obliged persons using the project initiated in $2018$ to subscribe for an electrical meter in ENEO agency and to pay the consumption every month. In addition, in March $2022$ a union of teachers for public schools, so called "One has suffered too much" stopped teaching during more than a month after submitting several requests to Cameroon government regarding valuable problems related to their profession, which last for more than a decade (the majority of them have worked during many years without salaries, housing, career prospects,...). In Cameroon, there is a direct correlation between poverty and corruption and generalized poverty is one of the main social causes of corruption. Regardless of the scope of corruption, such acts undetermine the development of civil society and exacerbate poverty, especially when public resources that should be used to improve people's life are abused or mismanaged by public officials \cite{gnm}. Here, the characteristics of the corruption are: bribery, conflict of interest, nepotism, illicit financial flow as defined in \cite{28gnm}, electoral financing, financial leakage, embezzlement, tax evasion, budget process, high levels corruption (presidential and parliamentary elections), bureaucratic corruption, false account and fraud \cite{37gnm}. This phenomenon has plagued the whole components of the country, particularly: Judicial, Military, Financial, Health, Educational and other departments. Furthermore, Cameroon has topped twice the list of most corrupt countries in the world in $1998$-$1999$ \cite{35gnm}. Although the government has put in place some Institutions (French Acronym (CONAC), Supreme State Audit (SSA), National Anti-Corruption Commission (NACC), Inter Employer Grouping of Cameroon (GICAM), Business Coalition Against Corruption (BCAC) and Special Criminal Court (SCC)) to fight against corruption in both public and private sectors, perception of this phenomenon continues to affect people trust in public services and justice systems. For instance, the actual president of republic did forty years as head of state and several of his collaborators occupied the post of minister during at least twenty five years. These public servants are from eighty years old to ninety. For more details, we refer the readers to \cite{8gnm,2gnm,13gnm,17gnm,25gnm,29gnm,36gnm,38gnm}. In this country of central Africa, corruption lies in the field of dangerous infectious diseases which are described as a pandemic. For most citizens the anti-corruption campaign has been merely a wasted effort since the government does not take any measures against the corrupted and corrupters. There is a general feeling that corruption and in particular high-levels corruption has been institutionalized to maintaining the same person as "lifetime" head of state \cite{17gnm}. Furthermore, the report \cite{12gnm} suggested that more than $14.7$ billion FCFA (i.e., $\$ 26.73$ million) has been lost by the State as a result of mismanagement and corruption. In this report, it is established that the most corrupt public services in Cameroon are mainly the ministry of finance (treasury, taxes, customs, central and decentralized services, etc...), public security (roadside check, establishment of passports and national identity cards, police custody, etc...), ministry of public contracts (recruitment, appointment of employees, career prospects,...), ministry of justice (corruption of magistrates and slowness in rendering justice,...) and ministry of defense (appointment of staff,...) \cite{12gnm,13gnm}. In \cite{11gnm} it is established that more than $4700$ billion FCFA (approximately, US $\$ 8.1$ billion) for about $20\%$ of the Gross Domestic Product (GDP) has been embezzled in the last decades and stashed in foreign banks by many high ranking members of government. However, no deeply study has been carried out to estimate the exact amount embezzled in the Cameroon budget since its year of independence, $1960$. This situation widely delayed and diverted economic growth and social development. Thus, a serious combat against corruption should highly improve the poverty reduction process.\\

      In the literature \cite{10ca,19ca,1ejo,2ejo,encovid}, a wide set of scientists have developed efficient numerical techniques in an approximate solution of mathematical models of the dynamic of poverty, corruption and infectious diseases. This class of model lies in the field of complex systems of ordinary differential equations (ODEs). The difficulties in computing exact solutions to such types of complex systems are closely related to those in solving certain class of nonlinear partial differential equations (PDEs) \cite{en7,en3,en4,en16,en5,en6,en12}. For such classes of problems, several authors have analyzed a broad range of statistical and numerical methods in approximate solutions. For more details, the readers can consult the works discussed in \cite{10ca,en15,en9,16ca,en11,2ejo,en13,19ca,en10,6rza,en14,12ca,en8} and references therein. In this paper, we develop an efficient third-step second-order convergent explicit numerical approach for solving a system of nonlinear ODEs modeled by the dynamic of poverty and corruption. The proposed numerical scheme is stable for any value of the initial condition, second-order convergence, fast and more efficient than a large class of numerical techniques widely studied in the literature for solving similar problems \cite{10ca,16ca,1ejo,4nrn,en1,en2,encovid}. The numerical experiments are performed using the data coming from Cameroon, a country located in central Africa and where corruption can be compared to a "pathology" and poverty reached the half of poverty line \cite{42gnm,4ca}. We recall that this paper deals with a computed solution of the mathematical model on the dynamic of corruption and poverty. Particularly, we are interested in the following four items:
      \begin{description}
        \item[i.] Mathematical reformulation on the dynamic of corruption and poverty,
        \item[ii.] detailed description of the third-step second-order explicit numerical scheme applied to problem mentioned in item i),
        \item[iii.] stability analysis and error estimates of the developed approach,
        \item[iv.] a large set of numerical examples which validate the theoretical studies and also considers the particular case of corruption and poverty in Cameroon.
      \end{description}

      The remainder of the work is the following: Section $\ref{sec2}$ deals with the mathematical model on the dynamic of corruption and poverty.
      In Section $\ref{sec3}$, we develop the third-step explicit numerical technique in an approximate solution of the considered problem. The stability together with the error estimates of the new algorithm are deeply analyzed in Section $\ref{sec4}$ whereas a large class of numerical evidences are discussed in Section $\ref{sec5}$. Finally, Section $\ref{sec6}$ considers the general conclusions and presents future works.

      \section{Mathematical model on the dynamic of corruption and poverty}\label{sec2}
      This section deals with a nonlinear model describing the dynamic of poverty and corruption. The population is assumed to be homogeneous (so that the spatial variable is neglected) divided into five compartments as follows:
      \begin{itemize}
        \item $y_{1}(t)$: Susceptible persons: the people in this class are never involved in any corruption which could impact the economic growth of the country,
        \item $y_{2}(t)$: Corrupt people: persons who are convicted in corrupt practices and are able to convince a susceptible individual to become corrupt,
        \item $y_{3}(t)$: Poverty class: individuals whose monthly incomes are below the poverty line,
        \item $y_{4}(t)$: Prosecuted/imprisoned individuals: people who are involved in corrupt acts and jailed for a specific period of time. During this period, they can neither be involved in any corrupt practices nor influence others,
        \item $y_{5}(t)$: Honest persons: individuals who could never be corrupted even whether their life conditions become bad.
      \end{itemize}
      With the above tools, we should analyze a mathematical model on the dynamic of poverty and corruption described in \cite{ejo} by the following system of nonlinear ordinary differential equations
      \begin{equation}\label{1}
      \frac{y_{1}(t)}{dt}=\theta-\frac{1}{N}(\alpha_{1}y_{1}y_{2}+\alpha_{2}y_{1}y_{3})(t)-(\gamma+\sigma)y_{1}(t),\text{\,\,\,\,\,\,\,\,\,\,\,on\,\,\,\,\,\,\,}
      I=(t_{0},T),
      \end{equation}
       \begin{equation}\label{2}
      \frac{y_{2}(t)}{dt}=\frac{1}{N}(\alpha_{1}y_{1}y_{2}+r_{2}y_{2}y_{3})(t)-(\gamma+b_{1}+\tau+r_{1})y_{2}(t)+\rho(1-\mu)y_{4}(t),\text{\,\,\,\,\,\,\,\,\,\,\,on\,\,\,\,\,\,\,}
      I=(t_{0},T),
      \end{equation}
       \begin{equation}\label{3}
      \frac{y_{3}(t)}{dt}=\frac{1}{N}(\alpha_{2}y_{1}y_{3}-r_{2}y_{2}y_{3})(t)+r_{1}y_{2}(t)-(\gamma+b_{2})y_{3}(t),\text{\,\,\,\,\,\,\,\,\,\,\,on\,\,\,\,\,\,\,}
      I=(t_{0},T),
      \end{equation}
       \begin{equation}\label{4}
      \frac{y_{4}(t)}{dt}=\tau y_{2}(t)-[\rho(1-\mu)+\rho\mu+\gamma]y_{4}(t),\text{\,\,\,\,\,\,\,\,\,\,\,on\,\,\,\,\,\,\,} I=(t_{0},T),
      \end{equation}
       \begin{equation}\label{5}
      \frac{y_{5}(t)}{dt}=\sigma y_{1}(t)+b_{1}y_{2}(t)+b_{2}y_{3}(t)+\rho\mu y_{4}(t)-\gamma y_{5}(t),\text{\,\,\,\,\,\,\,\,\,\,\,on\,\,\,\,\,\,\,}I=(t_{0},T),
      \end{equation}
      subjects to initial condition
      \begin{equation}\label{6}
      y_{1}(t_{0})=y_{1}^{0},\text{\,\,\,}y_{2}(t_{0})=y_{2}^{0},\text{\,\,\,}y_{3}(t_{0})=y_{3}^{0},\text{\,\,\,}y_{4}(t_{0})=y_{4}^{0},
      \text{\,\,\,}y_{5}(t_{0})=y_{5}^{0},
      \end{equation}
       where $\theta$ and $\gamma$ are the recruitment rate into the susceptible population and the death removal rate respectively, $1/\rho$ is the average period prosecuted persons spend in prison, $\rho\mu$ represents the transition rate from imprisoned to honest, $\rho(1-\mu)$ designates the transition rate from jailed to corrupt, $p_{1}$ and $p_{2}$ indicate the corruption transmission probability per contact and the poverty transmission probability per contact respectively, $\beta_{1}$ and $\beta_{2}$ denote both effort rates against corruption and poverty respectively, $\alpha_{1}=p_{1}(1-\beta_{1})$ and $\alpha_{2}=p_{2}(1-\beta_{2})$ are the effective corruption contact rate and the effective poverty contact rate respectively, $r_{1}$ and $r_{2}$ represent the rates at which both corrupt individuals and poor persons become poor and corrupt respectively, $\tau$ means the proportion of corrupt individuals prosecuted and jailed, $b_{1}$, $b_{2}$ and $\sigma$ indicate the proportion of corrupt persons, poor individuals and susceptible people, respectively, that become honest persons and $N$ is the size of the population in the considered country (for example: Cameroon). Without loss of the generality, we assume in this work that for a fixed period of time, the number of births equals the number of deaths. This assumption suggests that $N$ is constant. Furthermore, $y_{i}^{0}$, $i=1,2,...,5$, denote the initial data. It is worth noticing to mention that when the function $y_{3}$ vanishes on the interval $I=(t_{0},T)$, the initial-value problem $(\ref{1})$-$(\ref{6})$ provides the only model for dynamic of corruption, which is defined as
       \begin{equation*}
      \frac{y_{1}(t)}{dt}=\theta-(\gamma+\sigma)y_{1}(t)-\frac{1}{N}\alpha_{1}y_{1}(t)y_{2}(t),\text{\,\,\,\,\,\,\,\,\,\,\,on\,\,\,\,\,\,\,}
      I,
      \end{equation*}
        \begin{equation*}
      \frac{y_{2}(t)}{dt}=-(\gamma+b_{1}+\tau+r_{1})y_{2}(t)+\frac{1}{N}\alpha_{1}y_{1}(t)y_{2}(t)+\rho(1-\mu)y_{4}(t),\text{\,\,\,\,\,\,\,\,\,\,\,
      on\,\,\,\,\,\,\,}
      I,
      \end{equation*}
      \begin{equation*}
      \frac{y_{4}(t)}{dt}=\tau y_{2}(t)-[\rho(1-\mu)+\rho\mu+\gamma]y_{4}(t),\text{\,\,\,\,\,\,\,\,\,\,\,on\,\,\,\,\,\,\,} I,
      \end{equation*}
      \begin{equation*}
      \frac{y_{5}(t)}{dt}=\sigma y_{1}(t)+b_{1}y_{2}(t)+\rho\mu y_{4}(t)-\gamma y_{5}(t),\text{\,\,\,\,\,\,\,\,\,\,\,on\,\,\,\,\,\,\,}I,
      \end{equation*}
      subjects to initial condition
       \begin{equation*}
      y_{1}(t_{0})=y_{1}^{0},\text{\,\,\,}y_{2}(t_{0})=y_{2}^{0},\text{\,\,\,}y_{4}(t_{0})=y_{4}^{0},\text{\,\,\,}y_{5}(t_{0})=y_{5}^{0}.
      \end{equation*}
      However, the aim of this paper is not to focus the study on the only model of corruption. We introduce the following functions and vector functions:
      \begin{eqnarray*}
        y&:& I\rightarrow \mathbb{R}^{5}\\
         & & t\mapsto \left(y_{1}(t),...,y_{5}(t)\right),
      \end{eqnarray*}
       and
        \begin{eqnarray*}
        F&:& I\times\mathbb{R}^{5}\rightarrow \mathbb{R}^{5}\\
         & & (t,y)\mapsto \left(F_{1}(t,y),...,F_{5}(t,y)\right),
      \end{eqnarray*}
       where for $i=1,2,...,5$, $F_{i}:$ $I\times\mathbb{R}^{5}\rightarrow \mathbb{R}$ are defined as
       \begin{equation}\label{9}
        F_{1}(t,y)=\theta-\frac{1}{N}(\alpha_{1}y_{1}y_{2}+\alpha_{2}y_{1}y_{3})(t)-(\gamma+\sigma)y_{1}(t),
       \end{equation}
       \begin{equation}\label{10}
        F_{2}(t,y)=\frac{1}{N}(\alpha_{1}y_{1}y_{2}+r_{2}y_{2}y_{3})(t)-(\gamma+b_{1}+\tau+r_{1})y_{2}(t)+\rho(1-\mu)y_{4}(t),
       \end{equation}
       \begin{equation}\label{11}
        F_{3}(t,y)=\frac{1}{N}(\alpha_{2}y_{1}y_{3}-r_{2}y_{2}y_{3})(t)+r_{1}y_{2}(t)-(\gamma+b_{2})y_{3}(t),
       \end{equation}
       \begin{equation}\label{12}
        F_{4}(t,y)=\tau y_{2}(t)-[\rho(1-\mu)+\rho\mu+\gamma]y_{4}(t),
       \end{equation}
       \begin{equation}\label{13}
        F_{5}(t,y)=\sigma y_{1}(t)+b_{1}y_{2}(t)+b_{2}y_{3}(t)+\rho\mu y_{4}(t)-\gamma y_{5}(t).
       \end{equation}
       Plugging equations $(\ref{1})$-$(\ref{13})$ to obtain the following system of ODEs
      \begin{equation}\label{14}
      \frac{dy}{dt}=F(t,y), \text{\,\,\,\,\,\,\,\,\,\,\,on\,\,\,\,\,\,\,} I=(t_{0},T),
      \end{equation}
      with initial condition
      \begin{equation}\label{15}
      y(t_{0})=y^{0},
      \end{equation}
      where $\frac{dy}{dt}=(\frac{dy_{1}}{dt},...,\frac{dy_{5}}{dt})$ and $y^{0}=(y^{0}_{1},...,y^{0}_{5})$.\\

      Now, it follows from equations $(\ref{9})$-$(\ref{13})$ that the functions $F_{i}$ and their partial derivatives are continuously differentiable on the strip $\mathcal{S}=\{(t,y),\text{\,}t\in[t_{0},T],\text{\,}y\in\mathbb{R}^{5}\}$, whereas the partial derivatives $\partial y_{j}F_{i}$ are not bounded on this domain. By the Henrici result \cite{en22}, the initial-value problem $(\ref{1})$-$(\ref{6})$ has a unique solution $y(t)$ defined in a certain neighborhood $I(t_{0})\subset [t_{0},T]$ of the initial datum $t_{0}$. Since the system of nonlinear equations $(\ref{1})$-$(\ref{6})$ models a real-world problem and the size of the population $N=\underset{j=1}{\overset{5}\sum}y_{j}(t)$ is constant for any $t\in[t_{0},T]$, thus the functions $\partial y_{j}F_{i}$ are bounded on the compact domain $[t_{0},T]\times[0,N]^{5}$. So, it comes from the theory of differential equations \cite{en22} that the initial-value problem $(\ref{14})$-$(\ref{15})$ admits a unique solution $y(t)$ defined on the entire interval $[t_{0},T]$. This shows the existence and uniqueness of the analytical solution of the system of nonlinear equations $(\ref{1})$-$(\ref{6})$ or equivalently $(\ref{14})$-$(\ref{15})$.

       \section{Development of the three-step explicit numerical technique}\label{sec3}
       This section considers the construction of the third-step explicit numerical scheme in an approximate solution of the initial-value problem $(\ref{1})$-$(\ref{6})$, which describes the dynamic of corruption and poverty.\\

       Let $M$ be a positive integer. Set $k:=\Delta t=\frac{T-t_{0}}{M}$, be the steplength, $t_{n}=t_{0}+nk,$ and       $t_{n-\frac{1}{2}}=\frac{t_{n}+t_{n-1}}{2}$ for $n=1,2,...,M$. Consider $I_{k}=\{t_{n},0\leq n\leq M\}$, be a regular partition of $[t_{0},T]$ and let $\mathcal{F}_{k}=\{y^{n}=(y^{n}_{1},y^{n}_{2},y^{n}_{3},y^{n}_{4},y^{n}_{5}),\text{\,\,}n=0,1,..., M\},$ be the mesh functions space defined on
       \begin{equation}\label{15a}
        I_{k}\times\mathbb{R}^{5}\subset\mathcal{S}.
       \end{equation}
       We denote $y(t_{n})=y^{n}=(y^{n}_{1},...,y^{n}_{5})$ be the exact solution of the initial-value problem $(\ref{1})$-$(\ref{6})$ at the grid point $t_{n}$, while $Y^{n}=(Y^{n}_{1},...,Y^{n}_{5})$ stands for the solution provided by the proposed third-step explicit numerical procedure at the mesh point $t_{n}$.\\

        Furthermore, we introduce the following discrete norms
       \begin{equation}\label{16}
        \|y^{n}\|_{\infty}=\underset{1\leq i\leq5}{\max}|y_{i}^{n}|\text{\,\,\,and\,\,\,}\||y^{M}|\|_{L^{2}(I)}=
        \left(k\underset{n=1}{\overset{M}\sum}\|y^{n}\|_{\infty}^{2}\right)^{\frac{1}{2}},
       \end{equation}
       where $|\cdot|$ denotes the norm of the field of complex numbers $\mathbb{C}$.\\

       Expanding the Taylor series for the vector function $y$ about the grid point $t_{n}$ with step size $\frac{k}{6}$ using both forward and backward difference representations gives
       \begin{equation}\label{17}
       y^{n+\frac{1}{3}}=y^{n+\frac{1}{6}}+\frac{k}{6}y_{t}^{n+\frac{1}{6}}+\frac{k^{2}}{72}y_{2t}^{n+\frac{1}{6}}+\mathcal{O}(k^{3}),\text{\,\,\,\,}
       y^{n}=y^{n+\frac{1}{6}}-\frac{k}{6}y_{t}^{n+\frac{1}{6}}+\frac{k^{2}}{72}y_{2t}^{n+\frac{1}{6}}+\mathcal{O}(k^{3}),
       \end{equation}
       \begin{equation}\label{19}
       y^{n+\frac{2}{3}}=y^{n+\frac{1}{2}}+\frac{k}{6}y_{t}^{n+\frac{1}{2}}+\frac{k^{2}}{72}y_{2t}^{n+\frac{1}{2}}+\mathcal{O}(k^{3}),\text{\,\,\,\,}
       y^{n+\frac{1}{3}}=y^{n+\frac{1}{2}}-\frac{k}{6}y_{t}^{n+\frac{1}{2}}+\frac{k^{2}}{72}y_{2t}^{n+\frac{1}{2}}+\mathcal{O}(k^{3}),
       \end{equation}
       \begin{equation}\label{21}
        y^{n+1}=y^{n+\frac{5}{6}}+\frac{k}{6}y_{t}^{n+\frac{5}{6}}+\frac{k^{2}}{72}y_{2t}^{n+\frac{5}{6}}+\mathcal{O}(k^{3}),\text{\,\,\,\,}
       y^{n+\frac{2}{3}}=y^{n+\frac{5}{6}}-\frac{k}{6}y_{t}^{n+\frac{5}{6}}+\frac{k^{2}}{72}y_{2t}^{n+\frac{5}{6}}+\mathcal{O}(k^{3}),
       \end{equation}
       where $\mathcal{O}(k^{3})=(O(k^{3}),...,O(k^{3}))$ and $z_{mt}$ denotes the derivative of a vector function $z$ of order $m$. Subtracting the second equations in $(\ref{17})$, $(\ref{19})$ and $(\ref{21})$ from the first ones  and using equation $(\ref{14})$ provides
       \begin{equation*}
       y^{n+\frac{1}{3}}=y^{n}+\frac{k}{3}F(t_{n+\frac{1}{6}},y^{n+\frac{1}{6}})+\mathcal{O}(k^{3}),\text{\,\,\,\,\,\,}
       y^{n+\frac{2}{3}}=y^{n+\frac{1}{3}}+\frac{k}{3}F(t_{n+\frac{1}{2}},y^{n+\frac{1}{2}})+\mathcal{O}(k^{3}),
       \end{equation*}
       \begin{equation*}
        y^{n+1}=y^{n+\frac{2}{3}}+\frac{k}{3}F(t_{n+\frac{5}{6}},y^{n+\frac{5}{6}})+\mathcal{O}(k^{3}),
       \end{equation*}
       which are equivalent to
        \begin{equation}\label{26}
       y_{i}^{n+\frac{1}{3}}=y_{i}^{n}+\frac{k}{3}F_{i}(t_{n+\frac{1}{6}},y^{n+\frac{1}{6}})+O(k^{3}),
       \end{equation}
       \begin{equation}\label{27}
       y_{i}^{n+\frac{2}{3}}=y_{i}^{n+\frac{1}{3}}+\frac{k}{3}F_{i}(t_{n+\frac{1}{2}},y^{n+\frac{1}{2}})+O(k^{3}),
       \end{equation}
       \begin{equation}\label{28}
        y_{i}^{n+1}=y_{i}^{n+\frac{2}{3}}+\frac{k}{3}F_{i}(t_{n+\frac{5}{6}},y^{n+\frac{5}{6}})+O(k^{3}),
       \end{equation}
       for $i=1,2,...,5$. To construct the first-step of the desired numerical technique, we should approximate the term $F_{i}(t_{n+\frac{1}{6}},y^{n+\frac{1}{6}})$ by the sum
       \begin{equation}\label{29}
        c_{1}F_{i}(t_{n},y^{n})+c_{2}F_{i}(t_{n}+q_{1}k,y^{n}+q_{2}kF(t_{n},y^{n})),
       \end{equation}
       in such a way that the coefficients $c_{1}$, $c_{2}$, $q_{1}$ and $q_{2}$ are real numbers and chosen so that the Taylor series expansion results in
       \begin{equation*}
        \frac{y_{i}^{n+\frac{1}{3}}-y_{i}^{n}}{k/3}-F_{i}(t_{n+\frac{1}{6}},y^{n+\frac{1}{6}})=O(k^{2}).
       \end{equation*}
       The application of the Taylor series for $F_{i}$ and $y_{i}$ about the points $(t_{n},y^{n})$ and $t_{n}$ respectively, using forward difference formulation gives
       \begin{equation}\label{30}
        F_{i}(t_{n}+q_{1}k,y^{n}+q_{2}kF(t_{n},y^{n}))=F_{i}(t_{n},y^{n})+q_{1}k\partial_{t}F_{i}(t_{n},y^{n})+q_{2}k\underset{j=1}{\overset{5}\sum}\partial_{ y_{j}}F_{i}(t_{n},y^{n})+O(k^{2}),
       \end{equation}
       \begin{equation}\label{31}
        y_{i}^{n+\frac{1}{6}}=y_{i}^{n}+\frac{k}{3}y_{i,t}^{n}+\frac{k^{2}}{18}y_{i,2t}^{n}+O(k^{3})=y_{i}^{n}+\frac{k}{3}F_{i}(t_{n},y^{n})
        +\frac{k^{2}}{18}\frac{dF_{i}}{dt}(t_{n},y^{n})+O(k^{3}).
       \end{equation}
       But $\frac{dF_{i}}{dt}(t_{n},y^{n})=\partial_{t}F_{i}(t_{n},y^{n})+\underset{j=1}{\overset{5}\sum}F_{j}(t_{n},y^{n})\partial_{ y_{j}}F_{i}(t_{n},y^{n})$. This equation substituted into $(\ref{31})$ provides
       \begin{equation*}
        y_{i}^{n+\frac{1}{6}}=y_{i}^{n}+\frac{k}{3}F_{i}(t_{n},y^{n})+\frac{k^{2}}{18}\partial_{t}F_{i}(t_{n},y^{n})+\frac{k^{2}}{18}
        \underset{j=1}{\overset{5}\sum}F_{j}(t_{n},y^{n})\partial_{ y_{j}}F_{i}(t_{n},y^{n})+O(k^{3}),
       \end{equation*}
       which is equivalent to
       \begin{equation*}
        \frac{y_{i}^{n+\frac{1}{6}}-y_{i}^{n}}{k/3}=\left(F_{i}+\frac{k}{6}\partial_{t}F_{i}+\frac{k}{6}
        \underset{j=1}{\overset{5}\sum}F_{j}\partial_{ y_{j}}F_{i}\right)(t_{n},y^{n})+O(k^{2}).
       \end{equation*}
       Plugging this together with equations $(\ref{29})$ and $(\ref{30})$, it is not hard to see that
       \begin{equation*}
        \frac{y_{i}^{n+\frac{1}{3}}-y_{i}^{n}}{k/3}-F_{i}(t_{n+\frac{1}{6}},y^{n+\frac{1}{6}})=\left\{(1-c_{1}-c_{2})F_{i}+\frac{k}{6}
        \left[(1-6c_{2}q_{1})\partial_{t}F_{i}+(1-6c_{2}q_{2})\underset{j=1}{\overset{5}\sum}F_{j}\partial_{ y_{j}}F_{i}\right]\right\}(t_{n},y^{n})
       \end{equation*}
       \begin{equation*}
        +O(k^{2}).
       \end{equation*}
       The right-hand side of this equation equals $O(k^{2})$ if and only if: $1-c_{1}-c_{2}=0$, $1-6c_{2}q_{1}=0$ and $1-6c_{2}q_{2}=0$. This system of three equations with four unknowns has many solutions. For example, we can take
       \begin{equation}\label{32}
        c_{1}=c_{2}=\frac{1}{2} \text{\,\,\,\,\,and\,\,\,\,\,} q_{1}=q_{2}=\frac{1}{3}.
       \end{equation}
       Combining $(\ref{26})$, $(\ref{29})$, $(\ref{32})$ and rearranging terms, this yields
       \begin{equation}\label{33}
       y_{i}^{n+\frac{1}{3}}=y_{i}^{n}-\frac{k}{6}\left[F_{i}(t_{n},y^{n})+F_{i}\left(t_{n}+\frac{k}{3},y^{n}+\frac{k}{3}F(t_{n},y^{n})\right)\right]+O(k^{3}),
       \text{\,\,\,\,\,for\,\,\,\,\,} i=1,2,...,5.
       \end{equation}
      Tracking the infinitesimal term $O(k^{3})$ and replacing $y_{i}$ with the approximate solution $Y_{i}$, to get
       \begin{equation}\label{34}
       Y_{i}^{n+\frac{1}{3}}=Y_{i}^{n}-\frac{k}{6}\left[F_{i}(t_{n},Y^{n})+F_{i}\left(t_{n}+\frac{k}{3},Y^{n}+\frac{k}{3}F(t_{n},Y^{n})\right)\right],
       \text{\,\,\,\,\,for\,\,\,\,\,} i=1,2,...,5.
       \end{equation}
        Relation $(\ref{34})$ represents the first-step of the new algorithm.\\

        In a similar manner, with the Taylor series expansion for the function $y$ about the mesh points $t_{n+\frac{1}{3}}$ and $t_{n+\frac{2}{3}}$, with steplength $\frac{k}{6}$ utilizing forward difference scheme, one easily shows that
        \begin{equation}\label{35}
       y_{i}^{n+\frac{2}{3}}=y_{i}^{n+\frac{1}{3}}-\frac{k}{6}\left[F_{i}(t_{n+\frac{1}{3}},y^{n+\frac{1}{3}})+F_{i}\left(t_{n+\frac{1}{3}}+\frac{k}{3},
       y^{n+\frac{1}{3}}+\frac{k}{3}F(t_{n+\frac{1}{3}},y^{n+\frac{1}{3}})\right)\right]+O(k^{3}),
       \end{equation}
       and
        \begin{equation}\label{36}
       y_{i}^{n+1}=y_{i}^{n+\frac{2}{3}}-\frac{k}{6}\left[F_{i}(t_{n+\frac{2}{3}},y^{n+\frac{2}{3}})+F_{i}\left(t_{n+\frac{2}{3}}+\frac{k}{3},y^{n+\frac{2}{3}}
       +\frac{k}{3}F(t_{n+\frac{2}{3}},y^{n+\frac{2}{3}})\right)\right]+O(k^{3}),
       \end{equation}
       for $i=1,2,...,5$. Truncating the error terms in both equations $(\ref{35})$ and $(\ref{36})$, and replacing the analytical solution $y_{i}$ by the numerical one $Y_{i}$, to obtain the second-step and the third-step of the proposed numerical method:
        \begin{equation}\label{37}
       Y_{i}^{n+\frac{2}{3}}=Y_{i}^{n+\frac{1}{3}}-\frac{k}{6}\left[F_{i}(t_{n+\frac{1}{3}},Y^{n+\frac{1}{3}})+F_{i}\left(t_{n+\frac{1}{3}}+\frac{k}{3},
       Y^{n+\frac{1}{3}}+\frac{k}{3}F(t_{n+\frac{1}{3}},Y^{n+\frac{1}{3}})\right)\right],
       \end{equation}
       and
        \begin{equation}\label{38}
       Y_{i}^{n+1}=Y_{i}^{n+\frac{2}{3}}-\frac{k}{6}\left[F_{i}(t_{n+\frac{2}{3}},Y^{n+\frac{2}{3}})+F_{i}\left(t_{n+\frac{2}{3}}+\frac{k}{3},Y^{n+\frac{2}{3}}
       +\frac{k}{3}F(t_{n+\frac{2}{3}},Y^{n+\frac{2}{3}})\right)\right],
       \end{equation}
       for $i=1,2,...,5$. To start the algorithm, we should set
       \begin{equation}\label{39}
        Y^{0}=y^{0}.
       \end{equation}
       An assembly of equations $(\ref{34})$ and $(\ref{37})$-$(\ref{39})$ provides the new three-step explicit numerical approach for solving a mathematical model on the dynamic of corruption and poverty given by equations $(\ref{1})$-$(\ref{6})$, that is, for $n=0,1,2,...,M-1$, and $i=1,2,...,5$,
       \begin{equation}\label{s1}
       Y_{i}^{n+\frac{1}{3}}=Y_{i}^{n}-\frac{k}{6}\left[F_{i}(t_{n},Y^{n})+F_{i}\left(t_{n}+\frac{k}{3},Y^{n}+\frac{k}{3}F(t_{n},Y^{n})\right)\right],
       \end{equation}
        \begin{equation}\label{s2}
       Y_{i}^{n+\frac{2}{3}}=Y_{i}^{n+\frac{1}{3}}-\frac{k}{6}\left[F_{i}(t_{n+\frac{1}{3}},Y^{n+\frac{1}{3}})+F_{i}\left(t_{n+\frac{1}{3}}+\frac{k}{3},
       Y^{n+\frac{1}{3}}+\frac{k}{3}F(t_{n+\frac{1}{3}},Y^{n+\frac{1}{3}})\right)\right],
       \end{equation}
        \begin{equation}\label{s3}
       Y_{i}^{n+1}=Y_{i}^{n+\frac{2}{3}}-\frac{k}{6}\left[F_{i}(t_{n+\frac{2}{3}},Y^{n+\frac{2}{3}})+F_{i}\left(t_{n+\frac{2}{3}}+\frac{k}{3},Y^{n+\frac{2}{3}}
       +\frac{k}{3}F(t_{n+\frac{2}{3}},Y^{n+\frac{2}{3}})\right)\right],
       \end{equation}
       subjects to initial condition
       \begin{equation}\label{s4}
        Y_{i}^{0}=y_{i}^{0}.
       \end{equation}
       A combination of approximations $(\ref{s1})$-$(\ref{s4})$ yields an equivalent system of nonlinear equations, that is, for $n=0,1,2,...,M-1$, $i=1,2,...,5$,
        \begin{equation*}
       Y_{i}^{n+1}=Y_{i}^{n}-\frac{k}{6}\left[F_{i}(t_{n},Y^{n})+F_{i}(t_{n+\frac{1}{3}},Y^{n+\frac{1}{3}})+F_{i}(t_{n+\frac{2}{3}},Y^{n+\frac{2}{3}})
       +F_{i}\left(t_{n}+\frac{k}{3},Y^{n}+\frac{k}{3}F(t_{n},Y^{n})\right)+\right.
       \end{equation*}
        \begin{equation}\label{s5}
       \left.F_{i}\left(t_{n+\frac{1}{3}}+\frac{k}{3},Y^{n+\frac{1}{3}}+\frac{k}{3}F(t_{n+\frac{1}{3}},Y^{n+\frac{1}{3}})\right)
       +F_{i}\left(t_{n+\frac{2}{3}}+\frac{k}{3},Y^{n+\frac{2}{3}}+\frac{k}{3}F(t_{n+\frac{2}{3}},Y^{n+\frac{2}{3}})\right)\right],
       \end{equation}
       with initial condition
       \begin{equation*}
        Y_{i}^{0}=y_{i}^{0},\text{\,\,\,\,\,}Y_{i}^{\frac{1}{3}}=Y_{i}^{0}-\frac{k}{6}\left[F_{i}(t_{0},Y^{0})+F_{i}\left(t_{0}+\frac{k}{3},Y^{0}+\frac{k}{3}
        F(t_{0},Y^{0})\right)\right],
       \end{equation*}
        \begin{equation}\label{s6}
        Y_{i}^{\frac{2}{3}}=Y_{i}^{\frac{1}{3}}-\frac{k}{6}\left[F_{i}(t_{\frac{1}{3}},Y^{\frac{1}{3}})+F_{i}\left(t_{\frac{1}{3}}+\frac{k}{3},
       Y^{\frac{1}{3}}+\frac{k}{3}F(t_{\frac{1}{3}},Y^{\frac{1}{3}})\right)\right].
       \end{equation}
       It is worth mentioning that a three-step numerical method is widely encountered in the literature in the form given by formulation $(\ref{s5})$-$(\ref{s6})$. Furthermore, the first characteristic polynomial in term of $\lambda^{\frac{1}{3}}$ of the proposed three-step explicit numerical scheme is defined as
       \begin{equation*}
        P_{3}(\lambda^{\frac{1}{3}})=(\lambda^{\frac{1}{3}})^{3}-1=(\lambda^{\frac{1}{3}}-1)((\lambda^{\frac{1}{3}})^{2}+\lambda^{\frac{1}{3}}+1).
       \end{equation*}
       It is not hard to observe that the roots of this polynomial are: $\lambda_{1}^{\frac{1}{3}}=1$, $\lambda_{2}^{\frac{1}{3}}=-\frac{1}{2}-\widehat{i}\frac{\sqrt{3}}{2}$ and $\lambda_{3}^{\frac{1}{3}}=-\frac{1}{2}+\widehat{i}\frac{\sqrt{3}}{2}$, where $\widehat{i}$ is the complex number satisfying $\widehat{i}^{2}=-1$. Since the three roots lie in the closed unit disc and they are simple, it comes from the definition of zero-stability \cite{dahl1956,dahl1963} that the developed scheme $(\ref{s5})$-$(\ref{s6})$ is zero-stable. However, we should prove in this work that the new three-step explicit algorithm is stable with second-order convergence.

      \section{Analysis of stability and error estimates of the proposed technique}\label{sec4}
      In this section we analyze the stability and the convergence rate of the developed new approach $(\ref{s1})$-$(\ref{s4})$ applied to the initial-value      problem $(\ref{1})$-$(\ref{6}).$\\

      For the convenience of writing, we denote $\Delta_{i}(t_{m},y_{i}^{m})$ and $\widetilde{\Delta}_{i}(t_{m},Y_{i}^{m})$ be the difference
      quotients of the exact solution $y_{i}^{m}$ and the approximate one $Y_{i}^{m}$, respectively, at time $t_{m}$, for $m=n,n+\frac{1}{3},n+\frac{2}{3}$. More precisely, $\Delta_{i}(t_{m},y_{i}^{m})$ and $\widetilde{\Delta}_{i}(t_{m},Y_{i}^{m})$ are defined as
      \begin{equation}\label{40}
        \Delta_{i}(t_{m},y_{i}^{m})=\left\{
                                         \begin{array}{ll}
                                           \frac{y_{i}^{m+\frac{1}{3}}-y_{i}^{m}}{k/3}, & \hbox{if $k\neq0,$} \\
                                           \text{\,}\\
                                           F_{i}(t_{m},y_{i}^{m}), & \hbox{for $k=0$,}
                                         \end{array}
                                       \right.
      \end{equation}
       and
      \begin{equation}\label{41}
        \widetilde{\Delta}_{i}(t_{m},Y_{i}^{m})=\left\{
                                         \begin{array}{ll}
                                           \frac{Y_{i}^{m+\frac{1}{3}}-Y_{i}^{m}}{k/3}, & \hbox{if $k\neq0,$} \\
                                            \text{\,}\\
                                           F_{i}(t_{m},Y_{i}^{m}), & \hbox{for $k=0$,}
                                         \end{array}
                                       \right.
      \end{equation}
       for $m\in\{n,n+\frac{1}{3},n+\frac{2}{3}\}$ and $i=1,2,...,5$. The local discretization error term at the grid point
       $(t_{m},Y_{i}^{m})$ of the proposed numerical formulation $(\ref{s1})$-$(\ref{s4})$ is given by
       \begin{equation}\label{42}
        \delta_{i}(t_{m},y_{i}^{m})=\Delta_{i}(t_{m},Y_{i}^{m})-\widetilde{\Delta}_{i}(t_{m},Y_{i}^{m}),
        \text{\,\,\,}m=n,\text{\,\,}n+\frac{1}{3},\text{\,\,}n+\frac{2}{3}, \text{\,\,\,for\,\,\,}i=1,2,...,5,
       \end{equation}
        shows how the analytical solution of the initial-value problem $(\ref{1})$-$(\ref{6})$ efficiently approximates the solution provided by the developed third-step explicit numerical technique $(\ref{s1})$-$(\ref{s4})$.\\

        The following Lemma plays an important role when proving the stability and the convergence order of the new algorithm $(\ref{s1})$-$(\ref{s4})$.

       \begin{lemma}\cite{encovid,enarxivcovid}\label{l1}
       Let $v_{l}$ be a vector in $\mathbb{C}^{q}$ satisfying the estimates
       \begin{equation}\label{43}
        \|v_{m+1}\|_{L^{\infty}(\mathbb{C}^{q})}\leq(1+\psi_{1})\|v_{m}\|_{L^{\infty}(\mathbb{C}^{q})}+\psi_{2},,
       \end{equation}
       for $m=0,1,...,p$, where $p$ and $q$ are positive integers and $\psi_{1},\psi_{2}>0$ are two constants. Thus, it holds
        \begin{equation}\label{44}
        \|v_{m}\|_{L^{\infty}(\mathbb{C}^{q})}\leq e^{p\psi_{1}}\|v_{0}\|_{L^{\infty}(\mathbb{C}^{q})}+\frac{e^{p\psi_{1}}-1}{\psi_{1}}\psi_{2}.
       \end{equation}
       \end{lemma}

      \begin{theorem}\label{t}(Analysis of stability and error estimates)\\
      Suppose $y^{m}$ be the analytical solution of the system of nonlinear equations $(\ref{14})$-$(\ref{15})$ at time $t_{m}$ and let $Y^{m}$ be the solution provided by the proposed three-step numerical method $(\ref{s1})$-$(\ref{s4})$, or equivalently $(\ref{s5})$-$(\ref{s6})$, at time level $m$.    Let $e^{m}=y^{m}-Y^{m}$ be the global discretization error term at time $t_{m}$. Thus, the following inequalities are satisfied
       \begin{equation}\label{45}
      \||Y^{M}|\|_{L^{2}(I)}\leq \||y^{M}|\|_{L^{2}(I)}+\exp\left((T-t_{0})\widetilde{C}_{1}\right)\|e^{0}\|_{\infty}+\widetilde{C}_{2}
      \left(\exp((T-t_{0})\widetilde{C}_{1})-1\right)k^{2},
      \end{equation}
      and
       \begin{equation}\label{46}
      \||e^{M}|\|_{L^{2}(I)}\leq \widetilde{C}_{2}\left(\exp((T-t_{0})\widetilde{C}_{1})-1\right)k^{2},
      \end{equation}
      for every positive integer $M$, where $\widetilde{C}_{1}$ and $\widetilde{C}_{2}$ are two positive constants that do not depend on the step size $k$ and $\||\cdot|\|_{L^{2}(I)}$ is the norm defined by equation $(\ref{16})$.
      \end{theorem}
      It's worth mentioning that estimates $(\ref{45})$ and $(\ref{46})$ suggest that the developed numerical approach $(\ref{s1})$-$(\ref{s4})$ is stable for any value of the initial datum $Y^{0}$ and second-order accurate, respectively.

       \begin{proof}
       Firstly, we showed in Section $\ref{sec2}$ that the initial-value problem $(\ref{14})$-$(\ref{15})$ admits a unique solution $y$, defined on the interval $[t_{0},T]$. Let $\lambda_{0}$ be a positive constant independent of the grid size $k$.  We introduce the following domains:
       \begin{equation}\label{47}
       D_{i}=\{(t,x):\text{\,\,}t_{0}\leq t\leq T,\text{\,\,}x\in\mathbb{R},\text{\,\,}x-\lambda_{0}\leq y_{i}(t)\leq x+\lambda_{0}\},\text{\,\,\,}
       \mathcal{S}_{0}=\{(t,x):\text{\,\,}t_{0}\leq t\leq T,\text{\,\,}x\in(-\infty,\infty)\},
       \end{equation}
       where $y(t)=(y_{1}(t),...,y_{5}(t))$ is the exact solution of the system of differential equations $(\ref{14})$-$(\ref{15})$. A combination of equations $(\ref{s1})$-$(\ref{s3})$ and $(\ref{41})$, after rearranging terms results in
         \begin{equation}\label{48}
        \widetilde{\Delta}_{i}(t_{m},Y_{i}^{m})=-\frac{1}{2}\left[F_{i}(t_{m},Y^{m})+F_{i}\left(t_{m}+\frac{k}{3},Y^{m}+\frac{k}{3}
        F(t_{m},Y^{m})\right)\right],
        \end{equation}
        for $m\in\{n,n+\frac{1}{3},n+\frac{2}{3}\}$. Subtracting $(\ref{s1})$, $(\ref{s2})$ and $(\ref{s3})$, from $(\ref{33})$, $(\ref{35})$ and $(\ref{36})$, respectively, utilizing equations $(\ref{40})$-$(\ref{42})$, $(\ref{48})$ and rearranging terms, it is not hard to see that
        \begin{equation}\label{49}
        \delta_{i}(t_{m},y_{i}^{m})=\Delta_{i}(t_{m},y_{i}^{m})-\widetilde{\Delta}_{i}(t_{m},Y_{i}^{m})+O(k^{2}).
        \end{equation}
        Square modulus of this equation yields
        \begin{equation}\label{50}
        |\delta_{i}(t_{m},y_{i}^{m})|\leq C_{i,1}k^{2},\text{\,\,\,\,for\,\,\,\,}i=1,2,...,5,
        \end{equation}
        where $C_{i,1}>0$, are constant independent of the mesh size $k$. Setting $C_{1}=\underset{1\leq i\leq5}{\max}C_{i,1}$, and using the $L^{\infty}$-norm given by $(\ref{16})$, estimate $(\ref{50})$ implies
        \begin{equation}\label{51}
        \|\delta(t_{m},y^{m})\|_{\infty}=\underset{1\leq i\leq5}{\max}|\delta_{i}(t_{m},y_{i}^{m})|\leq C_{1}k^{2},
        \end{equation}
        where $\delta(t_{m},y^{m})=(\delta_{1}(t_{m},y_{1}^{m}),...,\delta_{5}(t_{m},y_{5}^{m}))$, and $m=n,n+\frac{1}{3},n+\frac{2}{3}$. Since the functions $F_{i}$, for $i=1,2,...,5$, given by equations $(\ref{9})$-$(\ref{13})$ are continuously differentiable, so equation $(\ref{48})$ shows that the functions $\widetilde{\Delta}_{i}$ defined by $(\ref{41})$ and their partial derivatives are continuous on the compact subset $D_{i}$. Thus, the Mean-Value Theorem suggests that there is a constant $\widetilde{L}_{i}>0$, that does not depend on the steplength $k$, so that
          \begin{equation}\label{52}
           |\widetilde{\Delta}_{i}(t,x_{1})-\widetilde{\Delta}_{i}(t,x_{2})|\leq \widetilde{L}_{i}|x_{1}-x_{2}|,
          \end{equation}
         for any pairs $(t,x_{1}),(t,x_{2})\in D_{i}$.\\

         Now, using the domain $\mathcal{S}_{0}$ given in $(\ref{47})$, we introduce the functions $\widehat{\Delta}_{i}$ defined on the strip $\mathcal{S}_{0}$ as follows
           \begin{equation}\label{53}
          \widehat{\Delta}_{i}(t,x)=\left\{
                                      \begin{array}{ll}
                                        \widetilde{\Delta}_{i}(t,x), & \hbox{if $(t,x)\in D_{i}$,} \\
                                        \text{\,}\\
                                        \widetilde{\Delta}_{i}(t,y_{i}(t)-\lambda_{0}), & \hbox{if $t_{0}\leq t\leq T$ and $x<y_{i}(t)-\lambda_{0}$,}\\
                                        \text{\,}\\
                                        \widetilde{\Delta}_{i}(t,y_{i}(t)+\lambda_{0}), & \hbox{if $t_{0}\leq t\leq T$ and $x>y_{i}(t)+\lambda_{0}$,}
                                      \end{array}
                                    \right.
           \end{equation}
            where $y=(y_{1}(t),...,y_{5}(t))$ denotes the analytical solution of the initial-value problem $(\ref{1})$-$(\ref{6})$. Estimate $(\ref{52})$ indicates that the functions $\widehat{\Delta}_{i}$ defined by $(\ref{53})$ are continuous and satisfy the "Lipschitz condition" on the domain $D_{i}$. Moreover, these functions are continuous and satisfy the "Lipschitz condition" on the domain $\mathcal{S}_{0}$. However, the proof of this result is obtained by replacing $\widehat{\delta}_{i}$ with $\widehat{\Delta}_{i}$ in the proof established in \cite{encovid}, page $16$.\\

            Combining approximations $(\ref{41})$ and $(\ref{48})$, it is not difficult to see that the third-step explicit formulation $(\ref{s1})$-$(\ref{s4})$ is generated by the functions $\widehat{\Delta}_{i}$, for $i=1,2,...,5$. Let $\widehat{Y}_{i}^{m}$ be the computed solution obtained at time level $m$, for $m=n,n+\frac{1}{3},n+\frac{2}{3}$, by replacing $\widetilde{\Delta}_{i}$ by $\widehat{\Delta}_{i}$, in equations $(\ref{41})$ and $(\ref{48})$. Thus, the solution $\widehat{Y}_{i}^{m}$ satisfies
             \begin{equation}\label{54}
            \widehat{Y}_{i}^{m+\frac{1}{3}}=\widehat{Y}_{i}^{m}+\frac{k}{3}\widehat{\Delta}_{i}(t_{m},\widehat{Y}_{i}^{m}),\text{\,\,\,\,for\,\,\,\,\,} m=n,n+\frac{1}{3},n+\frac{2}{3}, \text{\,\,\,and\,\,\,\,\,} i=1,2,...,5.
            \end{equation}
             Furthermore, it follows from equation $(\ref{40})$
            \begin{equation}\label{55}
            y_{i}^{m+\frac{1}{3}}=y_{i}^{m}+\frac{k}{3}\Delta_{i}(t_{m},y_{i}^{m}),\text{\,\,\,\,for\,\,\,\,\,} m\in\{n,n+\frac{1}{3},n+\frac{2}{3}\}, \text{\,\,\,and\,\,\,\,} i=1,2,...,5.
            \end{equation}
            Set $\widehat{e}^{m}=y^{m}-\widehat{Y}^{m}$, be the error term provided by the proposed numerical method $(\ref{54})$. Subtracting equation $(\ref{54})$ from $(\ref{55})$, it is easy to see that
            \begin{equation*}
            \widehat{e}_{i}^{m+\frac{1}{3}}=\widehat{e}_{i}^{m}+\frac{k}{3}\left(\Delta_{i}(t_{m},y_{i}^{m})-\widehat{\Delta}_{i}(t_{m},\widehat{Y}_{i}^{m})
            \right),
            \end{equation*}
            which is equivalent to
            \begin{equation*}
            \widehat{e}_{i}^{m+\frac{1}{3}}=\widehat{e}_{i}^{m}+\frac{k}{3}\left[(\widehat{\Delta}_{i}(t_{m},y_{i}^{m})
             -\widehat{\Delta}_{i}(t_{m},\widehat{Y}_{i}^{m}))+(\Delta_{i}(t_{m},y_{i}^{m})-\widehat{\Delta}_{i}(t_{m},y_{i}^{m}))\right].
            \end{equation*}
            The modulus in both sides of this equation yields
             \begin{equation}\label{56}
            |\widehat{e}_{i}^{m+\frac{1}{3}}|\leq |\widehat{e}_{i}^{m}|+\frac{k}{3}\left[|\widehat{\Delta}_{i}(t_{m},y_{i}^{m})
             -\widehat{\Delta}_{i}(t_{m},\widehat{Y}_{i}^{m})|+|\Delta_{i}(t_{m},y_{i}^{m})-\widehat{\Delta}_{i}(t_{m},y_{i}^{m})|\right].
            \end{equation}
            Since $(t_{m},y_{i}^{m})\in D_{i}$, a combination of equations $(\ref{49})$ and $(\ref{53})$ results in
            \begin{equation}\label{57}
            \Delta_{i}(t_{m},y_{i}^{m})-\widehat{\Delta}_{i}(t_{m},y_{i}^{m})=\delta_{i}(t_{m},y_{i}^{m}).
            \end{equation}
            Utilizing the "Lipschitz condition" of the function $\widehat{\Delta}_{i}$ defined on the strip $\mathcal{S}_{0}$ by $(\ref{47})$ together with relations $(\ref{57})$ and $(\ref{51})$, estimate $(\ref{56})$ becomes
             \begin{equation}\label{58}
            |\widehat{e}_{i}^{m+\frac{1}{3}}|\leq |\widehat{e}_{i}^{m}|+\frac{Lk}{3}|y_{i}^{m}-\widehat{Y}_{i}^{m}|+\frac{C_{2}k^{3}}{3}\leq\left(1+\frac{Lk}{3}\right)|\widehat{e}_{i}^{m}|+
             \frac{C_{2}k^{3}}{3},
            \end{equation}
            for $m=n,n+\frac{1}{3},n+\frac{2}{3}$, $i=1,2,...,5$, where $L=\underset{1\leq i\leq5}{\max}\widetilde{L}_{i}$ and $C_{2}$ is a positive constant independent of the step size $k$. Taking the maximum in both sides of $(\ref{58})$ and using the $L^{\infty}$-norm given $(\ref{16})$ to get
             \begin{equation*}
            \|\widehat{e}^{m+\frac{1}{3}}\|_{\infty}\leq \left(1+\frac{Lk}{3}\right)\|\widehat{e}^{m}\|_{\infty}+\frac{C_{2}k^{3}}{3}.
            \end{equation*}
             This is equivalent to
              \begin{equation}\label{59}
            \|\widehat{e}^{n+\frac{1}{3}}\|_{\infty}\leq \left(1+\frac{Lk}{3}\right)\|\widehat{e}^{n}\|_{\infty}+\frac{C_{2}k^{3}}{3},
            \end{equation}
             \begin{equation}\label{60}
            \|\widehat{e}^{n+\frac{2}{3}}\|_{\infty}\leq \left(1+\frac{Lk}{3}\right)\|\widehat{e}^{n+\frac{1}{3}}\|_{\infty}+\frac{C_{2}k^{3}}{3},
            \end{equation}
             \begin{equation}\label{61}
            \|\widehat{e}^{n+1}\|_{\infty}\leq \left(1+\frac{Lk}{3}\right)\|\widehat{e}^{n+\frac{2}{3}}\|_{\infty}+\frac{C_{2}k^{3}}{3}.
            \end{equation}
            Substituting estimate $(\ref{59})$ into $(\ref{60})$ and the obtained result into $(\ref{61})$ to obtain
            \begin{equation}\label{62}
            \|\widehat{e}^{n+1}\|_{\infty}\leq \left(1+\frac{Lk}{3}\right)^{3}\|\widehat{e}^{n}\|_{\infty}+C_{2}\left(1+\frac{Lk}{9}\right)k^{3},
            \end{equation}
            for $n=0,1,2,...,M-1$. Since $\left(1+\frac{Lk}{3}\right)^{3}=1+Lk+(Lk)^{2}+\frac{1}{9}L^{3}k^{3}=1+kL_{1}$, where $L_{1}=L\left(1+Lk+\frac{1}{9}L^{2}k^{2}\right)$, it is easy to observe that inequality $(\ref{62})$ satisfies assumption $(\ref{43})$ of Lemma $\ref{l1}$. So, this estimate implies
            \begin{equation*}
            \|\widehat{e}^{n+1}\|_{\infty}\leq \exp(MkL_{1})\|\widehat{e}^{0}\|_{\infty}+\frac{L_{2}}{L_{1}}(\exp(MkL_{1})-1)k^{2},
            \end{equation*}
             where $L_{2}=C_{2}(1+\frac{1}{9}Lk)$. We remind that $L=\underset{1\leq i\leq5}{\max}\widetilde{L}_{i}$, where $\widetilde{L}_{i}$ denotes the constant of "Lipschitz condition" of the function $\widehat{\Delta}_{i}$ defined on $\mathcal{S}_{0}$ and $C_{2}$ is a positive constant independent of the mesh size $k$. Since $Mk=T-t_{0}$, it holds
             \begin{equation*}
            \|\widehat{e}^{n+1}\|_{\infty}\leq \exp((T-t_{0})L_{1})\|\widehat{e}^{0}\|_{\infty}+L_{2}L_{1}^{-1}(\exp((T-t_{0})L_{1})-1)k^{2}.
            \end{equation*}
           Taking the square in both sides of this inequality gives
           \begin{equation*}
            \|\widehat{e}^{n+1}\|_{\infty}^{2}\leq \left[\exp((T-t_{0})L_{1})\|\widehat{e}^{0}\|_{\infty}+L_{2}L_{1}^{-1}(\exp((T-t_{0})L_{1})-1)k^{2}\right]^{2}.
            \end{equation*}
            Summing this up from $n=0,1,2,...,M-1$, multiplying the obtained estimate by $k$ and using the definition of the norm $\||\cdot|\|_{L^{2}(I)}$
            given by $(\ref{16})$, to get
            \begin{equation*}
            \||\widehat{e}^{M}|\|_{L^{2}(I)}^{2}\leq \left[\exp((T-t_{0})L_{1})\|\widehat{e}^{0}\|_{\infty}+L_{2}L_{1}^{-1}(\exp((T-t_{0})L_{1})-1)k^{2}\right]^{2}.
            \end{equation*}
            The square root in both sides of this estimate results in
            \begin{equation}\label{63}
            \||\widehat{e}^{M}|\|_{L^{2}(I)}\leq \exp((T-t_{0})L_{1})\|\widehat{e}^{0}\|_{\infty}+L_{2}L_{1}^{-1}(\exp((T-t_{0})L_{1})-1)k^{2}.
            \end{equation}
            But $\||\widehat{Y}^{M}|\|_{L^{2}(I)}-\||y^{M}|\|_{L^{2}(I)}\leq\||\widehat{Y}^{M}-y^{M}|\|_{L^{2}(I)}=\||\widehat{e}^{M}|\|_{L^{2}(I)}$, this fact combined with $(\ref{63})$ imply
            \begin{equation}\label{64}
            \||\widehat{Y}^{M}|\|_{L^{2}(I)}\leq \||y^{M}|\|_{L^{2}(I)}+\exp((T-t_{0})L_{1})\|\widehat{e}^{0}\|_{\infty}
            +L_{2}L_{1}^{-1}(\exp((T-t_{0})L_{1})-1)k^{2}.
            \end{equation}
            Since $y$ is the analytical solution of the initial-value problem $(\ref{14})$-$(\ref{15})$, then $y$ is bounded on the closed interval $[t_{0},T]$. This estimate indicates that the numerical solutions $\widehat{Y}_{i}^{n}$, generated by the functions $\widehat{\Delta}_{i}$ are bounded. So, it comes from the definition of $\widehat{\Delta}_{i}$ given by relation $(\ref{53})$ that
            \begin{equation}\label{65}
            \widehat{Y}_{i}^{n}=Y_{i}^{n},\text{\,\,\,for\,\,\,}n=0,1,2,...,M\text{\,\,\,and\,\,\,}i=1,2,...,5.
            \end{equation}
            Thus
            \begin{equation*}
            \||Y^{M}|\|_{L^{2}(I)}\leq\||y^{M}|\|_{L^{2}(I)}+\exp((T-t_{0})L_{1})\|\widehat{e}^{0}\|_{\infty}+L_{2}L_{1}^{-1}(\exp((T-t_{0})L_{1})-1)k^{2},
            \end{equation*}
            for any positive integer $M$. This estimate shows that the proposed third-step numerical scheme $(\ref{s1})$-$(\ref{s3})$ is stable for any value of the initial datum $Y^{0}$ obtained from a one-step scheme such as the explicit Euler method. Now, it follows from the initial condition $(\ref{s4})$ and equation $(\ref{65})$ that $\|\widehat{e}^{0}\|_{\infty}=\|e^{0}\|_{\infty}=0$. This fact together with estimate $(\ref{63})$ and equation $(\ref{65})$ provides
            \begin{equation*}
            \||e^{M}|\|_{L^{2}(I)}\leq L_{2}L_{1}^{-1}(\exp((T-t_{0})L_{1})-1)k^{2},
            \end{equation*}
            for every positive integer $M$. This completes the proof of Theorem $\ref{t}$.
           \end{proof}
           The analysis presented in both Sections $\ref{sec3}$ and $\ref{sec4}$, suggests that the developed third-step second-order explicit numerical approach $(\ref{s1})$-$(\ref{s4})$ is fast and more efficient than a large class of numerical methods widely studied in the literature in an approximate solution of systems of nonlinear differential equations of type $(\ref{14})$-$(\ref{15})$. Indeed, the proposed algorithm $(\ref{s1})$-$(\ref{s4})$, is a third-step scheme (less consuming in a computer memory savings), explicit, unconditionally stable and second-order accurate (very fast). It also requires only one initial datum instead of two initial data like it is usually encountered in the literature.

         \section{Numerical experiments}\label{sec5}
         In this section, we carry out some numerical experiments and present computational results to show the stability and the accuracy of the new third-step second-order explicit technique $(\ref{s1})$-$(\ref{s4})$ applied to a mathematical model on the dynamic of corruption and poverty $(\ref{1})$-$(\ref{5})$ or equivalently $(\ref{14})$, with suitable initial condition $(\ref{6})$. We consider two examples given \cite{lgi} and we compute and plot the exact solution ($y^{n}$), the computed one ($Y^{n}$) and the error term ($e^{n}$), versus $n$ using formula $(\ref{16})$. In addition, we evaluate the convergence rate ($R(2k/k)$) of the new algorithm corresponding to the ratio of the approximation errors associated with two steplengths $2k$ and $k$:
         \begin{equation*}
            \||y^{M}|\|_{L^{2}(I)}=\left(k\underset{n=1}{\overset{M}\sum}\|y^{n}\|_{\infty}^{2}\right)^{\frac{1}{2}},\text{\,\,\,}
            \||Y^{M}|\|_{L^{2}(I)}=\left(k\underset{n=1}{\overset{M}\sum}\|Y^{n}\|_{\infty}^{2}\right)^{\frac{1}{2}},\text{\,\,\,}
            \||e^{M}|\|_{L^{2}(I)}=\left(k\underset{n=1}{\overset{M}\sum}\|y^{n}-Y^{n}\|_{\infty}^{2}\right)^{\frac{1}{2}},
         \end{equation*}
         \begin{equation*}
           R(2k/k)=log_{2}\left(\frac{\||e^{M}|\|_{L^{2}(I)}}{\||e^{2M}|\|_{L^{2}(I)}}\right).
         \end{equation*}
         Furthermore, we focus on the particular case of the dynamic of poverty and corruption in Cameroon, where some data are taken in \cite{28gnm,12gnm,35gnm,37gnm,36gnm,gnm,sf,cmt}. In these works, the authors conducted the analysis on poverty and corruption from its independence in $1960$ to date. Since the country's economic performance has been strongly unstable during the considered period $[1960,2022]$, this time interval should be subdivided into three subperiods: $[1960,1986[$, $[1986,2002[$ and $[2022,2022]$, in which each subinterval must summarizes the situation on the dynamic of corruption and poverty in Cameroon. This suggests that each parameter described in the system of nonlinear equations $(\ref{1})$-$(\ref{5})$ should take three values. Moreover: $N\in\{10^{7},1.6\times10^{7},2.5\times10^{7}\}$, $\theta\in\{0.2,0.3,0.25\}$, $\gamma\in\{0.2,0.3,0.25\}$, $1/\rho\in\{0.2,0.35,0.3\}$, $\rho\mu\in\{0.55,0.3,0.35\}$, $\rho(1-\mu)\in\{0.45,0.7,0.65\}$, $p_{1}\in\{0.3,0.8,0.75\}$, $p_{2}\in\{0.1,0.4,0.35\}$, $\alpha_{1}\in\{0.018,0.72,0.5625\}$, $\alpha_{2}\in\{0.03,0.34,0.228\}$, $r_{1}\in\{0.45,0.8,0.75\}$, $r_{2}\in\{0.5,0.9,0.8\}$, $\tau\in\{0.6,0.15,0.3\}$, $b_{1}\in\{0.3,0.10,0.15\}$, $b_{2}\in\{0.3,0.12,0.15\}$, $\sigma\in\{0.9,0.6,0.8\}$, $\beta_{1}\in\{0.6,0.10,0.25\}$ and $\beta_{2}\in\{0.7,0.15,0.40\}$. In addition, we set
         $y_{1}(t_{0})\in\{3.5\times10^{6},2.4\times10^{6},5\times10^{6}\}$, $y_{2}(t_{0})\in\{1.5\times10^{6},3.2\times10^{6},4.25\times10^{6}\}$,
         $y_{3}(t_{0})\in\{1.5\times10^{6},6.4\times10^{6},9.5\times10^{6}\}$, $y_{4}(t_{0})\in\{0.5\times10^{6},2.4\times10^{6},3\times10^{6}\}$
         and $y_{5}(t_{0})\in\{3\times10^{6},1.6\times10^{6},3.25\times10^{6}\}$. We recall that the first element in the set $\{a_{1},a_{2},a_{3}\}$ is $a_{1}$ and it is associated with the subperiod $[1960,1986[$, the second element in this set, $a_{2}$ is associated with the subinterval $[1986,2002[$ and the last element in this set denoted $a_{3}$ is related to the subperiod $[2002,2022]$. In each subperiod, we compute different values of $Y_{i}(t_{n})$, for $n=0,1,2,...,M$, $i\in\{1,2,...,5\}$ and we plot the approximate solution $Y_{i}$ versus $n$.\\

         $\bullet$ \textbf{Example $1.$} We test the efficiency of the new approach $(\ref{s1})$-$(\ref{s4})$ using the following system of nonlinear equations given in \cite{lgi} by
         \begin{eqnarray*}
           \frac{dy_{1}(t)}{dt} &=& -y_{1}(t)+g_{1}(t), \\
           \frac{dy_{2}(t)}{dt} &=& y_{1}(t)-y_{2}^{2}(t)+g_{2}(t), \text{\,\,\,\,\,\,\,\,\,\,\,\,\,\,\,\,\,\,\,\,\,\,on\,\,\,\,}I=(0,1),\\
           \frac{dy_{3}(t)}{dt} &=& y_{2}^{2}(t)+g_{3}(t),
         \end{eqnarray*}
         subjects to initial condition
         \begin{equation*}
         y_{1}(0)=y_{2}(0)=y_{3}(0)=0,
         \end{equation*}
         where
         \begin{equation*}
         g_{1}(t)=t^{2}+t-1;\text{\,\,\,\,\,}g_{2}(t)=(t^{2}-t)^{2}e^{-2t}-(t^{2}-3t+1)e^{-t}-t^{2}+t;\text{\,\,\,\,\,}g_{3}(t)=
         -(t^{2}-t)^{2}e^{-2t}+(t-1)\cos(t)+\sin(t).
         \end{equation*}
         The analytical solution is given in \cite{lgi} by: $y_{1}(t)=t^{2}-t$, $y_{2}(t)=(t^{2}-t)e^{-t}$, $y_{3}(t)=(t-1)\sin(t)$. \\

          \textbf{Table 1.}  $\label{T1}$. Stability and convergence accuracy $R(2k/k),$ of the new three-step explicit technique with varying steplength $k$.

           \begin{equation*}
         \begin{tabular}{|c|c|c|c|c|}
            \hline
            $k$ & $\||y^{M}|\|_{L^{2}(I)}$ &$\||Y^{M}|\|_{L^{2}(I)}$ & $\||E^{M}|\|_{L^{2}(I)}$ & $R(2k/k)$\\
            \hline
            $2^{-4}$ & $1.8235\times10^{-1}$  & $1.7712\times10^{-1}$  &  $7.3475\times10^{-3}$ &   --  \\
            \hline
            $2^{-5}$ & $1.8261\times10^{-1}$  & $1.7756\times10^{-1}$  &  $2.1106\times10^{-3}$ & 1.7987 \\
            \hline
            $2^{-6}$ & $1.8260\times10^{-1}$  & $1.7953\times10^{-1}$  &  $5.3296\times10^{-4}$ & 1.9783\\
            \hline
            $2^{-7}$ & $1.8258\times10^{-1}$  & $1.8068\times10^{-1}$  &  $1.3296\times10^{-4}$ & 2.0029 \\
           \hline
            $2^{-8}$ & $1.8179\times10^{-1}$  & $1.8261\times10^{-1}$  &  $3.3238\times10^{-5}$ & 2.0001 \\
            \hline
          \end{tabular}
            \end{equation*}
           \text{\,}\\

         $\bullet$ \textbf{Example $2.$} Let $I=(0,1)$, we test the following initial-value problem defined in \cite{lgi} as
         \begin{eqnarray*}
           \frac{dy_{1}(t)}{dt} &=& -y_{1}(t)+y_{2}(t)y_{3}(t)+g_{1}(t), \\
           \frac{dy_{2}(t)}{dt} &=& y_{1}(t)-y_{2}(t)y_{3}(t)+g_{2}(t), \text{\,\,\,\,\,\,\,\,\,\,\,\,\,\,\,\,\,\,\,\,\,\,on\,\,\,\,}I=(0,1),\\
           \frac{dy_{3}(t)}{dt} &=& y_{2}^{2}(t)+g_{3}(t),
         \end{eqnarray*}
         subjects to initial condition
         \begin{equation*}
         y_{1}(0)=y_{2}(0)=y_{3}(0)=0,
         \end{equation*}
         where
         \begin{equation*}
         g_{1}(t)=t^{2}+t-1-t(t-1)^{2}\sin(t),\text{\,\,\,\,\,}g_{2}(t)=t-t^{2}+(t^{2}-t)^{2}e^{-2t}-(t^{2}-3t+1)e^{-t}+t(t-1)^{2}e^{-t}\sin(t),
         \end{equation*}
         \begin{equation*}
         g_{3}(t)=-(t^{2}-t)^{2}e^{-2t}+(t-1)\cos(t)+\sin(t).
         \end{equation*}
         The analytical solution is given in \cite{lgi} by: $y_{1}(t)=t^{2}-t$, $y_{2}(t)=(t^{2}-t)e^{-t}$, $y_{3}(t)=(t-1)\sin(t)$.\\

          \textbf{Table 2.}  $\label{T2}$. Stability and convergence rate $R(2k/k),$ of the proposed three-step explicit approach with varying step size $k$.

           \begin{equation*}
         \begin{tabular}{|c|c|c|c|c|}
            \hline
            $k$ & $\||y^{M}|\|_{L^{2}(I)}$ &$\||Y^{M}|\|_{L^{2}(I)}$ & $\||E^{M}|\|_{L^{2}(I)}$ & $R(2k/k)$\\
            \hline
            $2^{-4}$ & $1.8209\times10^{-1}$  & $1.7782\times10^{-1}$  &  $8.0143\times10^{-3}$ &   --  \\
            \hline
            $2^{-5}$ & $1.8217\times10^{-1}$  & $1.7856\times10^{-1}$  &  $2.401\times10^{-3}$ & 1.7627 \\
            \hline
            $2^{-6}$ & $1.8235\times10^{-1}$  & $1.8079\times10^{-1}$  &  $6.4862\times10^{-4}$ & 1.8876\\
            \hline
            $2^{-7}$ & $1.8255\times10^{-1}$  & $1.8134\times10^{-1}$  &  $1.6943\times10^{-4}$ & 1.9367 \\
           \hline
            $2^{-8}$ & $1.8263\times10^{-1}$  & $1.8248\times10^{-1}$  &  $4.2566\times10^{-5}$ & 1.9929 \\
            \hline
          \end{tabular}
            \end{equation*}
           \text{\,}\\
           \text{\,}\\
           Both Tables $1$-$2$ and Figures $\ref{fig1}$-$\ref{fig2}$ suggest that the developed numerical technique $(\ref{s1})$-$(\ref{s4})$ is second-order accurate and stable for any initial datum $Y^{0}$ obtained from a one-step numerical scheme. This computational result confirms the theoretical one provided in Section $\ref{sec4}$, Theorem $\ref{t}$. Furthermore, the new algorithm described by $(\ref{s1})$-$(\ref{s4})$ is more faster and efficient than a wide set of numerical methods applied to systems of nonlinear differential equations \cite{4nrn,en1,en2,encovid,enarxivcovid,10ca,16ca}.\\

         Tables $3$-$5$ below present the situation on the dynamic of poverty and corruption in Cameroon using the new third-step second-order explicit technique $(\ref{s1})$-$(\ref{s4})$.\\

         $\bullet$ \textbf{Example $3.$} In this test, we set $t_{0}=1960$, $T=1986$, $N=10^{7}$, $k=10^{-3}$, $\theta=0.2$, $\gamma=0.2$, $1/\rho=0.2$, $\rho\mu=0.55$, $\rho(1-\mu)=0.45$, $p_{1}=0.3$, $p_{2}=0.1$, $\alpha_{1}=0.018$, $\alpha_{2}=0.03$, $r_{1}=0.45$, $r_{2}=0.5$, $\tau=0.6$, $b_{1}=0.3$, $b_{2}=0.3$, $\sigma=0.9$, $\beta_{1}=0.6$ and $\beta_{2}=0.7$. The initial condition is: $y_{1}(t_{0})=3.5\times10^{6}$,  $y_{2}(t_{0})=1.5\times10^{6}$, $y_{3}(t_{0})=1.5\times10^{6}$, $y_{4}(t_{0})=0.5\times10^{6}$ and $y_{5}(t_{0})=3\times10^{6}$.\\

         \textbf{Table 3.} Susceptible persons, corrupted people, poor individuals, prosecuted/jailed people and honest persons in Cameroon from $1960$ to $1985$.
         \begin{equation*}
           \small{\begin{tabular}{|c|c|c|c|c|c|c|c|}
            \hline
             People  & $[1960,1965[$            & $[1965,1970[$        & $[1970,1975[$        & $[1975,1980[$        & $[1980,1986[$ & Average &Rate\\
            \hline
             $y_{1}$ & $3.5077\times10^{6}$ & $3.5233\times10^{6}$ & $3.5428\times10^{6}$ & $3.5664\times10^{6}$ & $3.5862\times10^{6}$ & $3.5453\times10^{6}$  & $35.5\%$\\
            \hline
            $y_{2}$ & $1.5040\times10^{6}$ & $1.5119\times10^{6}$ & $1.5219\times10^{6}$ & $1.5340\times10^{6}$ & $1.5442\times10^{6}$ & $1.5225\times10^{6}$ & $15\%$\\
            \hline
            $y_{3}$ & $1.5003\times10^{6}$ & $1.5010\times10^{6}$ & $1.5019\times10^{6}$ & $1.5028\times10^{6}$ & $1.5036\times10^{6}$ & $1.5019\times10^{6}$ & $15\%$\\
            \hline
            $y_{4}$ & $0.4994\times10^{6}$ & $0.4982\times10^{6}$ & $0.4966\times10^{6}$ & $0.4947\times10^{6}$ & $0.4930\times10^{6}$ & $0.4964\times10^{6}$ & $5\%$\\
            \hline
            $y_{5}$ & $2.9938\times10^{6}$ & $2.9813\times10^{6}$ & $2.9656\times10^{6}$ & $2.9467\times10^{6}$ & $2.9308\times10^{6}$ & $2.9636\times10^{6}$ & $30\%$\\
            \hline
          \end{tabular}}
         \end{equation*}
           \text{\,}\\

            $\bullet$ \textbf{Example $4.$} In this example, $t_{0}=1986$, $T=2002$, $N=1.6\times10^{7}$, $k=10^{-3}$, $\theta=0.3$, $\gamma=0.3$, $1/\rho=0.35$, $\rho\mu=0.3$, $\rho(1-\mu)=0.7$, $p_{1}=0.8$, $p_{2}=0.4$, $\alpha_{1}=0.72$, $\alpha_{2}=0.34$, $r_{1}=0.8$, $r_{2}=0.9$, $\tau=0.15$, $b_{1}=0.1$, $b_{2}=0.12$, $\sigma=0.6$, $\beta_{1}=0.1$ and $\beta_{2}=0.15$. The initial condition is: $y_{1}(t_{0})=2.4\times10^{6}$,  $y_{2}(t_{0})=3.2\times10^{6}$, $y_{3}(t_{0})=6.4\times10^{6}$, $y_{4}(t_{0})=2.4\times10^{6}$ and $y_{5}(t_{0})=1.6\times10^{6}$.\\

         \textbf{Table 4.} Susceptible individuals, corrupted persons, poor people, prosecuted/imprisoned persons and honest people in Cameroon from $1986$ to $2001$

         \begin{equation*}
           \begin{tabular}{|c|c|c|c|c|c|c|}
            \hline
             People  & $[1986,1990[$            & $[1990,1994[$        & $[1994,1998[$        & $[1998,2002[$         & Average & Rate \\
            \hline
            $y_{1}$ & $2.4191\times10^{6}$ & $2.4411\times10^{6}$ & $2.4634\times10^{6}$ & $2.4859\times10^{6}$ &  $2.4524\times10^{6}$ & $15.3\%$\\
            \hline
            $y_{2}$ & $3.2080\times10^{6}$ & $3.2170\times10^{6}$ & $3.2259\times10^{6}$ & $3.2347\times10^{6}$ &  $3.2214\times10^{6}$ & $20.1\%$\\
            \hline
            $y_{3}$ & $6.4067\times10^{6}$ & $6.4143\times10^{6}$ & $6.4218\times10^{6}$ & $6.4294\times10^{6}$ &  $6.4181\times10^{6}$ & $40.1\%$\\
            \hline
            $y_{4}$ & $2.4186\times10^{6}$ & $2.4400\times10^{6}$ & $2.4616\times10^{6}$ & $2.4834\times10^{6}$ &  $2.4509\times10^{6}$ & $15.3\%$\\
            \hline
            $y_{5}$& $1.5838\times10^{6}$ & $1.5651\times10^{6}$ & $1.5462\times10^{6}$ & $1.5272\times10^{6}$ &  $1.5556\times10^{6}$ & $9.7\%$\\
            \hline
          \end{tabular}
         \end{equation*}
           \text{\,}\\

            $\bullet$ \textbf{Example $5.$} In this test, we take $t_{0}=2002$, $T=2022$, $N=2.5\times10^{7}$, $k=10^{-3}$, $\theta=0.25$, $\gamma=0.25$, $1/\rho=0.3$, $\rho\mu=0.35$, $\rho(1-\mu)=0.65$, $p_{1}=0.75$, $p_{2}=0.38$, $\alpha_{1}=0.563$, $\alpha_{2}=0.228$, $r_{1}=0.75$, $r_{2}=0.8$, $\tau=0.3$, $b_{1}=0.15$, $b_{2}=0.15$, $\sigma=0.8$, $\beta_{1}=0.25$ and $\beta_{2}=0.4$, with initial condition: $y_{1}(t_{0})=5\times10^{6}$,  $y_{2}(t_{0})=4.25\times10^{6}$, $y_{3}(t_{0})=9.5\times10^{6}$, $y_{4}(t_{0})=3\times10^{6}$ and $y_{5}(t_{0})=3.25\times10^{6}$.\\

          \textbf{Table 5.} Susceptible people, corrupted individuals, poor persons, prosecuted/jailed people and honest individuals in Cameroon from $2002$ to $2022$
           \begin{equation*}
           \small{\begin{tabular}{|c|c|c|c|c|c|c|c|}
            \hline
             People  & $[2002,2006[$            & $[2006,2010[$        & $[2010,2014[$        & $[2014,2018[$        & $[2018,2022[$ & Average &Rate\\
            \hline
            $y_{1}$ & $5.0301\times10^{6}$ & $5.0726\times10^{6}$ & $5.1094\times10^{6}$ & $5.1526\times10^{6}$ & $5.1899\times10^{6}$ & $5.1109\times10^{6}$  & $20.4\%$\\
            \hline
            $y_{2}$ & $4.2622\times10^{6}$ & $4.2793\times10^{6}$ & $4.2940\times10^{6}$ & $4.3111\times10^{6}$ & $4.3258\times10^{6}$ & $4.2945\times10^{6}$ & $17.2\%$\\
            \hline
            $y_{3}$ & $9.5073\times10^{6}$ & $9.5176\times10^{6}$ & $9.5263\times10^{6}$ & $9.5365\times10^{6}$ & $9.5452\times10^{6}$ & $9.5266\times10^{6}$ & $38.1\%$\\
            \hline
            $y_{4}$ & $3.0124\times10^{6}$ & $3.0299\times10^{6}$ & $3.0449\times10^{6}$ & $3.0626\times10^{6}$ & $3.0778\times10^{6}$ & $3.0455\times10^{6}$ & $12.2\%$\\
            \hline
            $y_{5}$ & $3.2237\times10^{6}$ & $3.1866\times10^{6}$ & $3.1545\times10^{6}$ & $3.1168\times10^{6}$ & $3.0843\times10^{6}$ & $3.1532\times10^{6}$ & $12.6\%$\\
            \hline
          \end{tabular}}
         \end{equation*}
           \text{\,}\\
           It is worth noticing to mention that the number of corrupted people equals $y_{2}+y_{4}$ whereas the number of poor persons is $y_{3}$. Furthermore, the computational results provided in both rows two and four of Tables $3$-$5$ only consider the high-levels corruption as defined in Section $\ref{sec1}$.\\

           From this study, we observe from Tables $3$-$5$ and Figures $\ref{fig3}$-$\ref{fig5}$ that corruption and poverty are less rampant in Cameroon during the period from $1960$ to $1986$. Specifically, an average of $20\%$ of civil servants are involved in corrupt acts and approximately $15\%$ of citizens are poor, whereas interval $[1986,2002[$ indicates that $40.1\%$ of cameroonians have incomes below the poverty line and an average of $35.4\%$ of civil officials are involved in corruption. In addition, the results provided in the subperiod $[2002,2022]$ shows that $38.1\%$ of people have expenditures below the poverty line and more than $29.4\%$ of persons working in public sector are still interested in corrupt practices. These numerical results confirm a large class of studies carried out in the literature on the considered country \cite{cmt,28gnm,35gnm,37gnm,17gnm,38gnm,36gnm,sf}. Indeed, the worst situation observed in the subperiod $[1986,2002[$ is partly due to the fact that, after the devaluation of the currency (F CFA) in $1990$, the economy witnessed a serious economic crisis and the poverty rate grew. In $1993$, the salaries of civil servants have been twice reduced (January and November), which caused their incomes less than twice.
           
           \subsection*{Some efficient strategies to fight against corruption and reduce poverty in\\ Cameroon}

           To overcome the worst situation of corruption and poverty in Cameroon, the government must implement the programme the president of republic has advertised during his electoral campaign. The electoral code must be revised in such a way that the electoral commission should be independent. Specifically, the ruling party and the first five opposition parties should agree on minimum requirements regarding the electoral code before the next presidential and parliamentary elections. This must avoid post-electoral crisis and stop the ruling party to be reelected indefinitely. There must be in each polling station one or two representative(s) of every political parties involved in the election. As regards the presidential, the president of the republic and the vice-president should be declared at the same time (French speaker for the first one and English speaker for the second one or vice-versa). The president of the constitutional court should freely declare president, vice-president or members of national assembly. Finally, citizens should know and participate in budget discussions ranging from the national assembly to the local councils. The suggested solutions should efficiently eradicate high-levels corruption which maintains the pretty one and thus, considerably reduces poverty in Cameroon.

         \section{General conclusion and future works}\label{sec6}
\begin{figure}
         \begin{center}
          Analysis of stability and convergence order of the new algorithm.
          \begin{tabular}{c c}
         \psfig{file=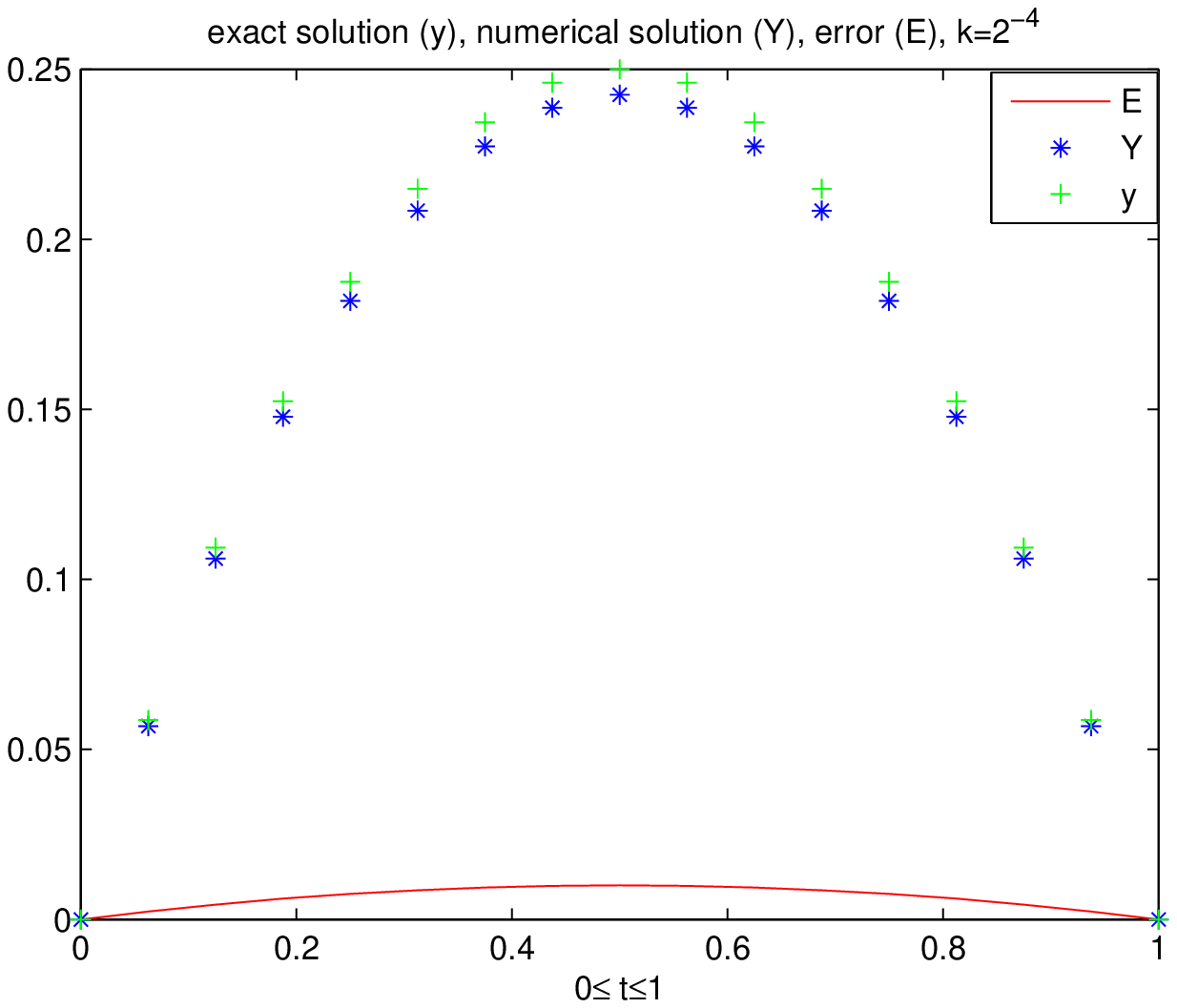,width=6cm}  & \psfig{file=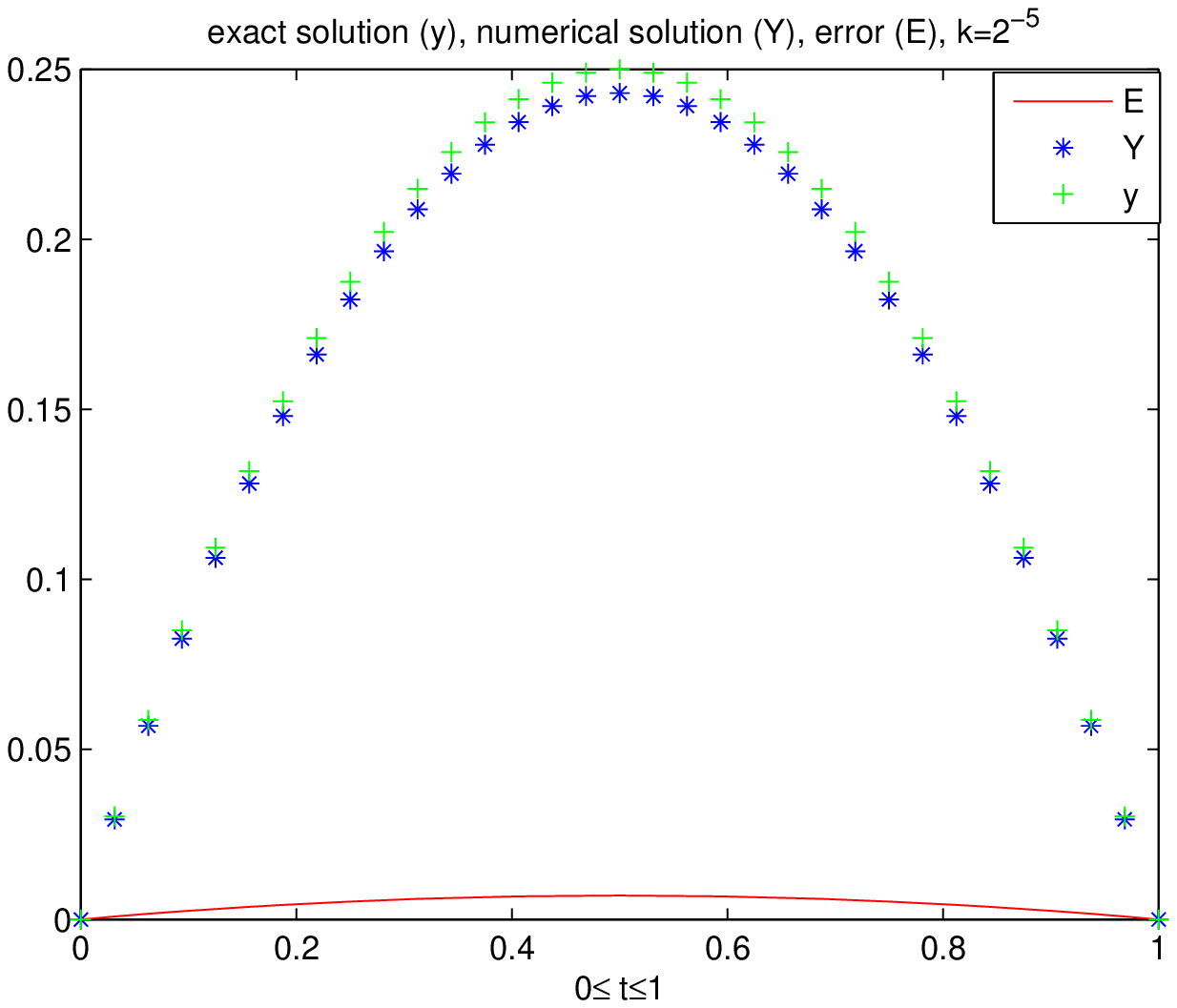,width=6cm}\\
         \psfig{file=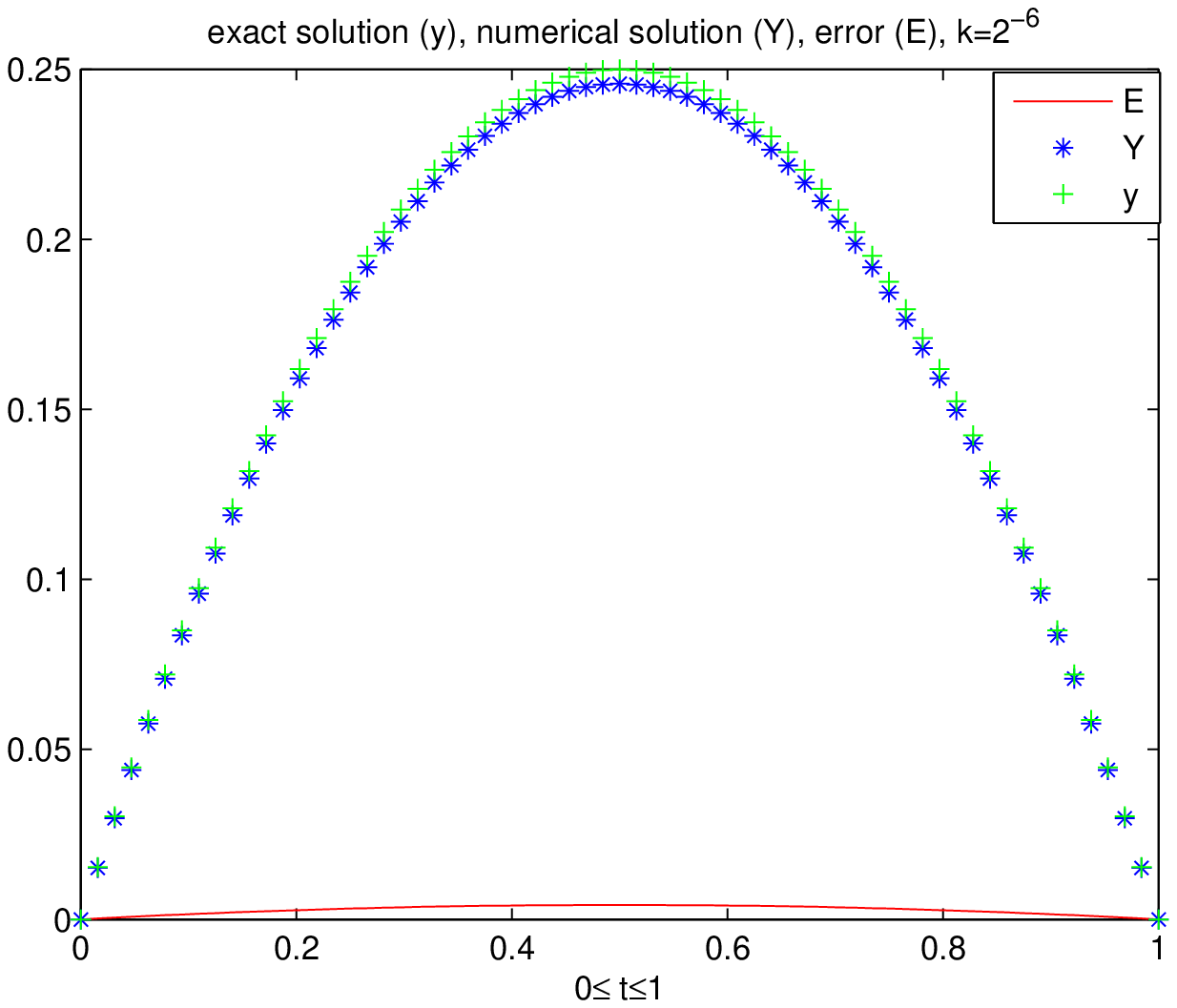,width=6cm}  & \psfig{file=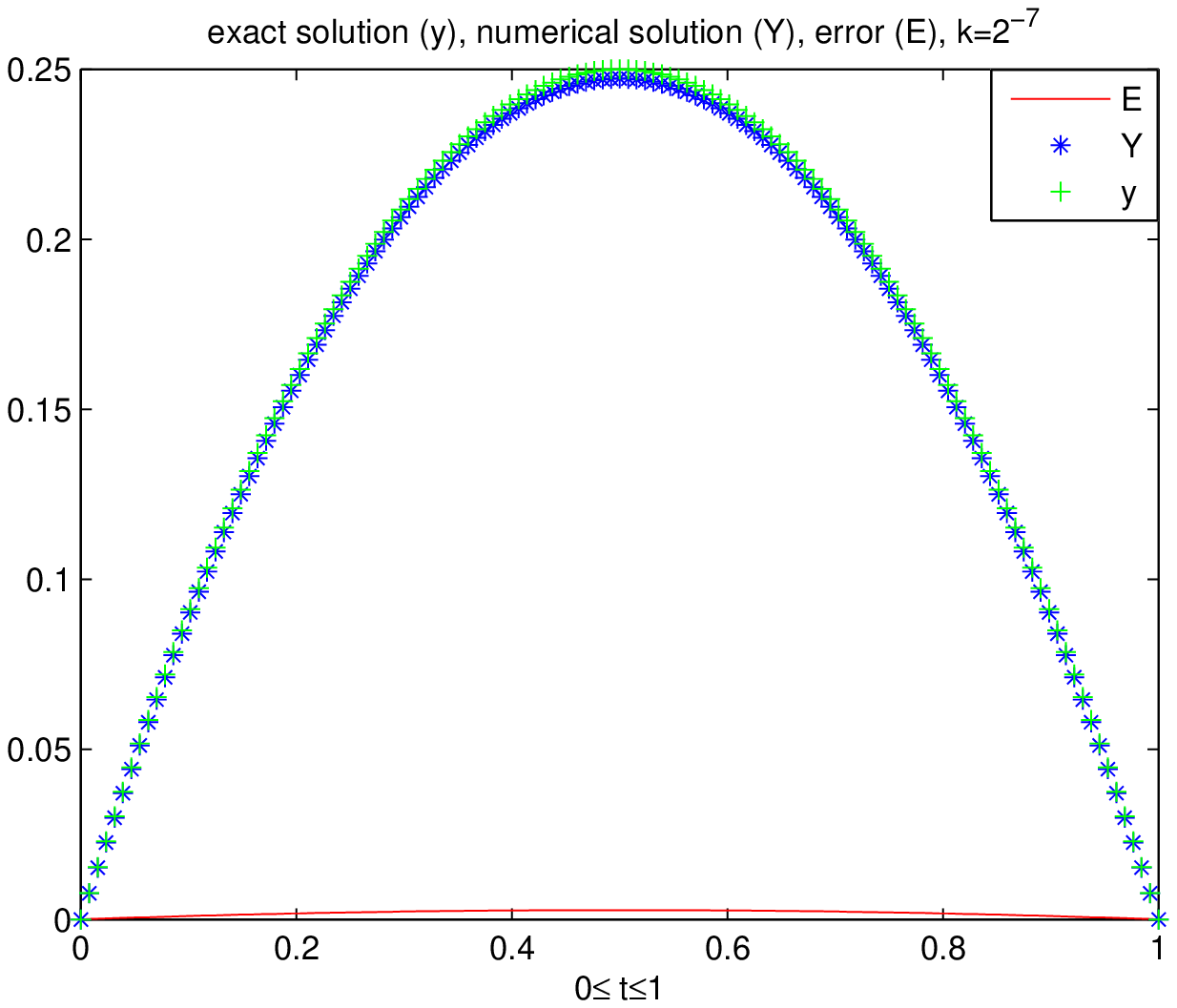,width=6cm}\\
         \end{tabular}
        \end{center}
         \caption{Exact solution (y: in green), Numerical solution (Y: in blue) and Error (E: in red) for Example 1}
        \label{fig1}
          \end{figure}

\begin{figure}
         \begin{center}
         Stability and convergence accuracy of the developed approach.
          \begin{tabular}{c c}
         \psfig{file=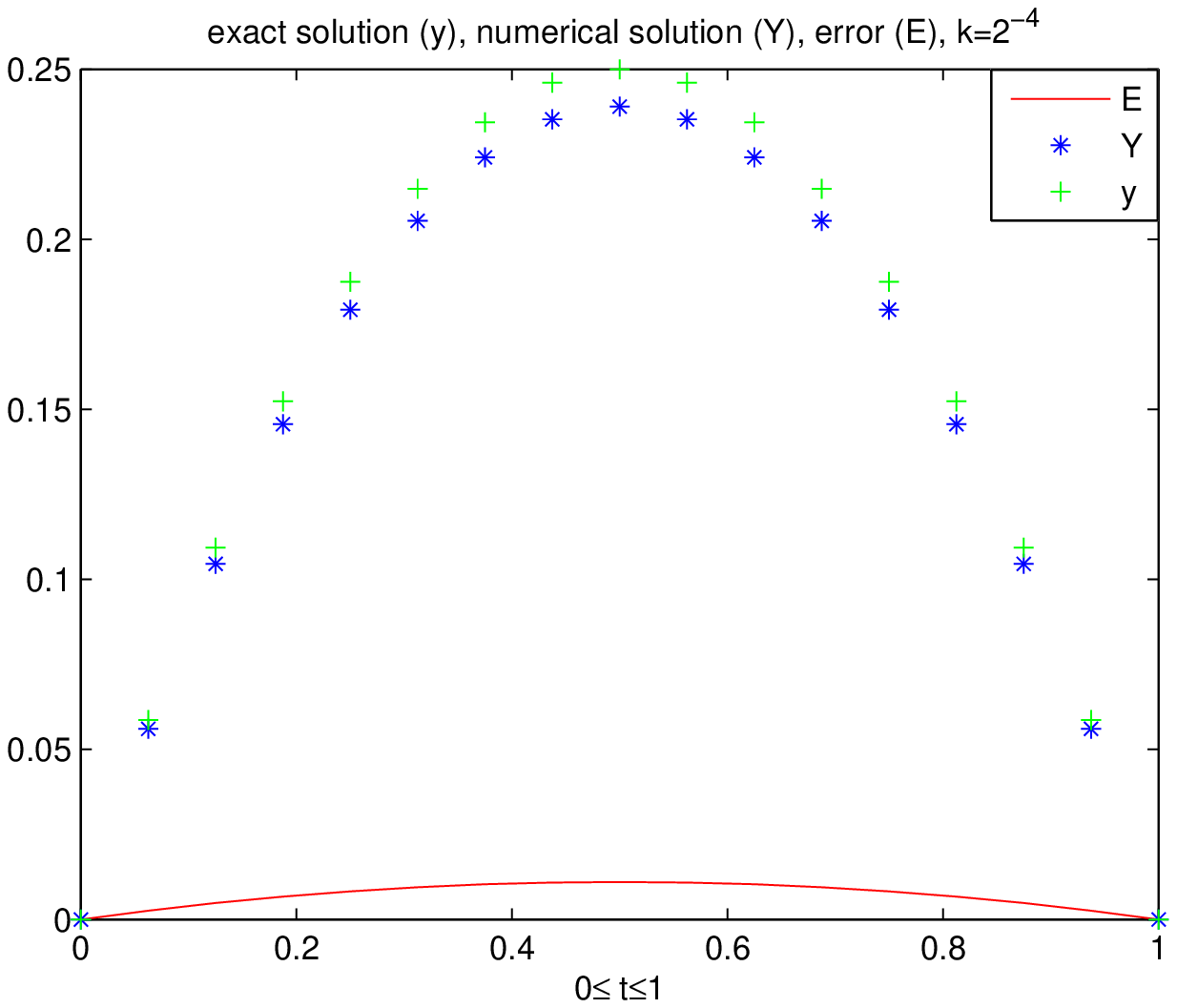,width=6cm}  & \psfig{file=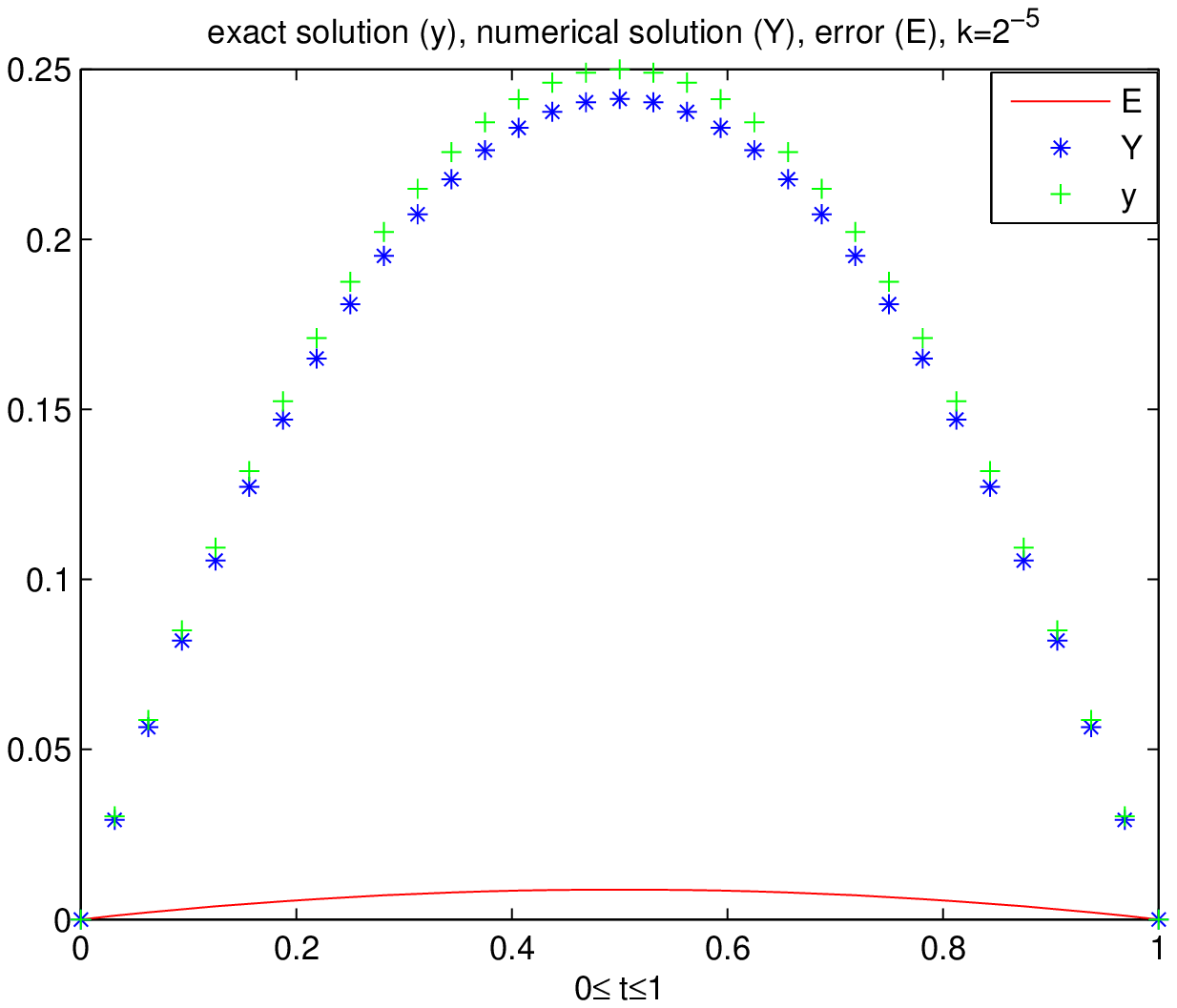,width=6cm}\\
         \psfig{file=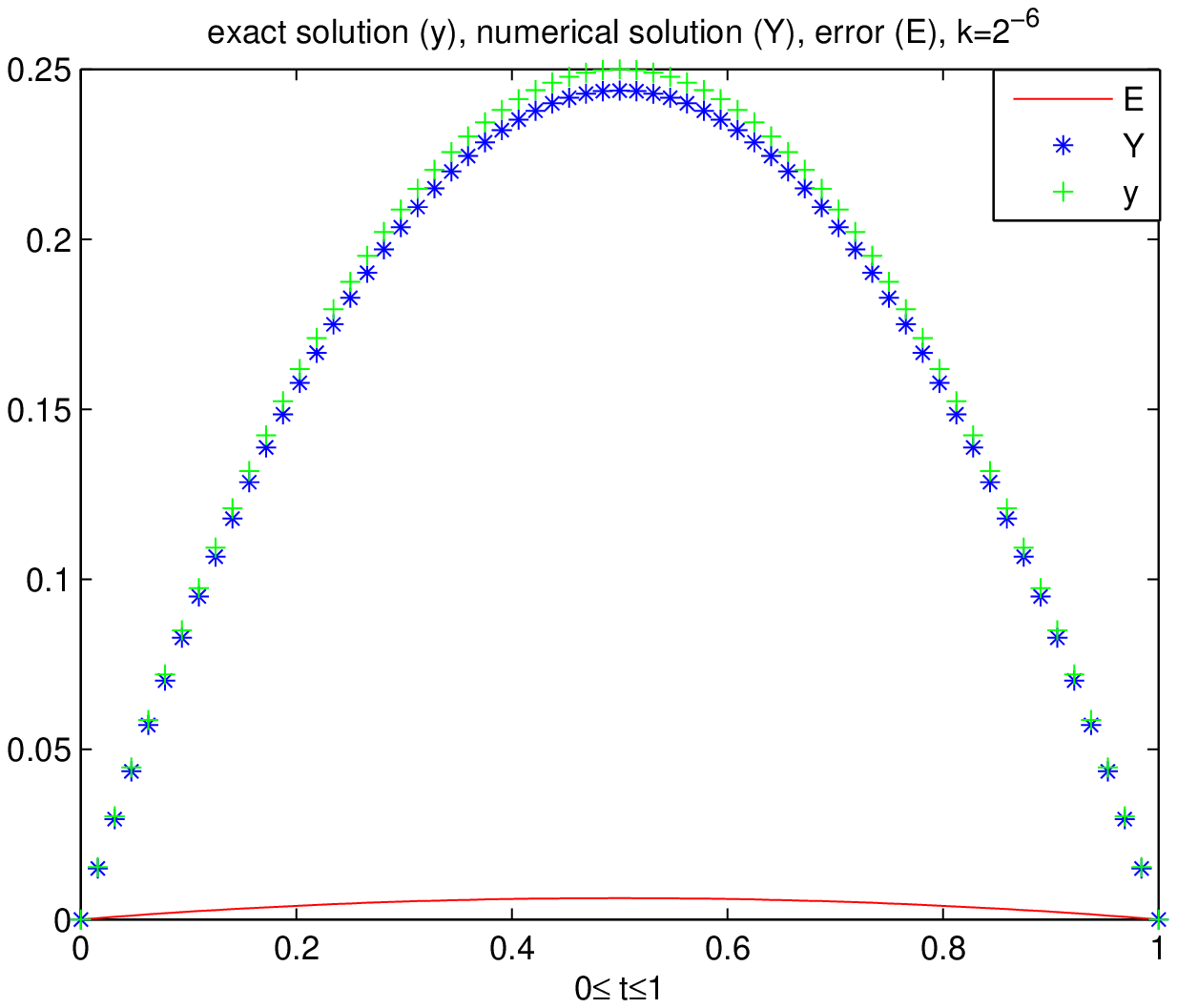,width=6cm}  & \psfig{file=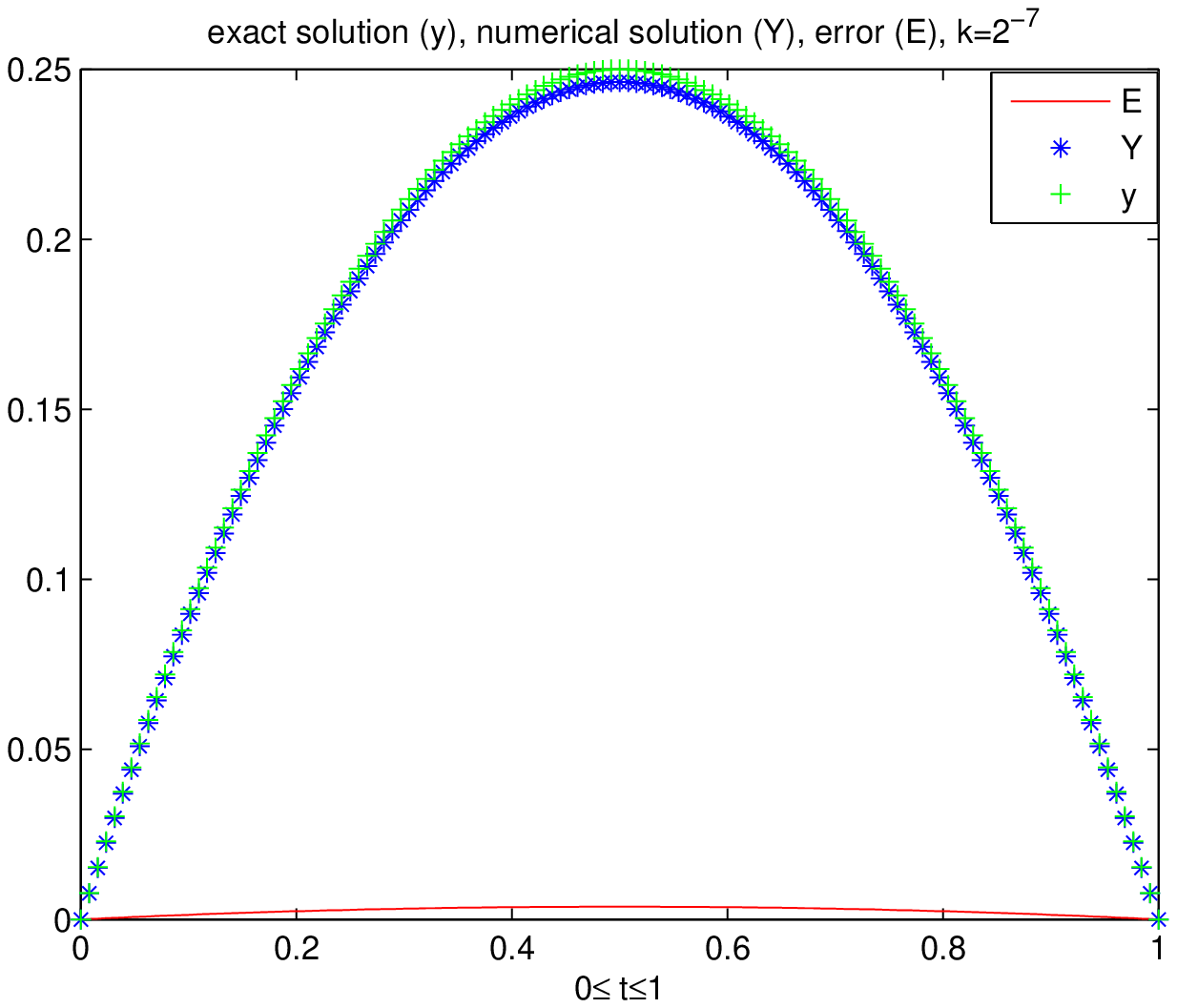,width=6cm}\\
         \end{tabular}
        \end{center}
        \caption{Analytical solution (y: in green), Computed solution (Y: in blue), Error (E: in red) for Example 2}
          \label{fig2}
          \end{figure}

\begin{figure}
         \begin{center}
         Analysis of corruption and poverty in Cameroon: $1960$-$1986$.
          \begin{tabular}{c}
         \psfig{file=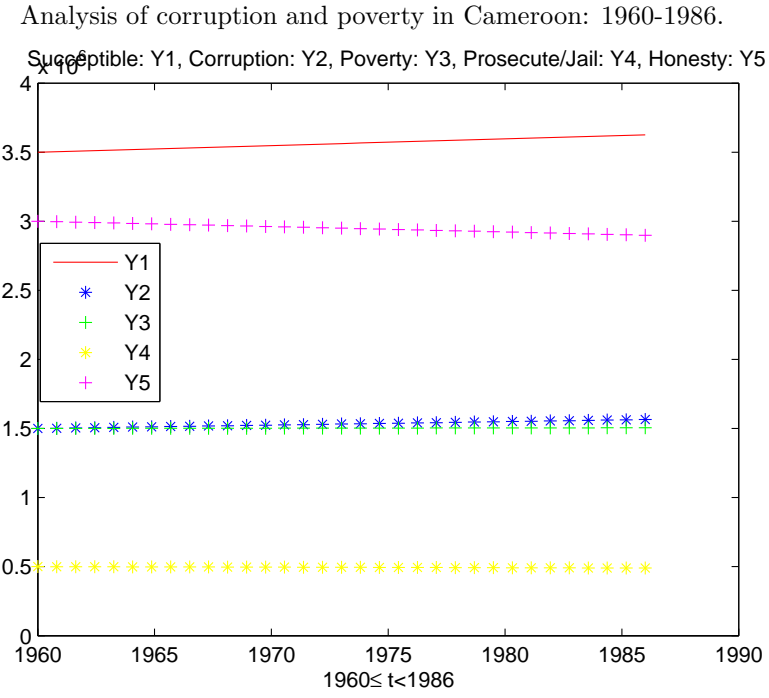,width=12cm}
         \end{tabular}
        \end{center}
        \caption{Exact solution (y: in green), Approximate solution (Y: in blue), Error (E: in red) for Example 3}
          \label{fig3}
          \end{figure}

\begin{figure}
         \begin{center}
         Situation of corruption and poverty in Cameroon: $1986$-$2002$.
          \begin{tabular}{c}
           \psfig{file=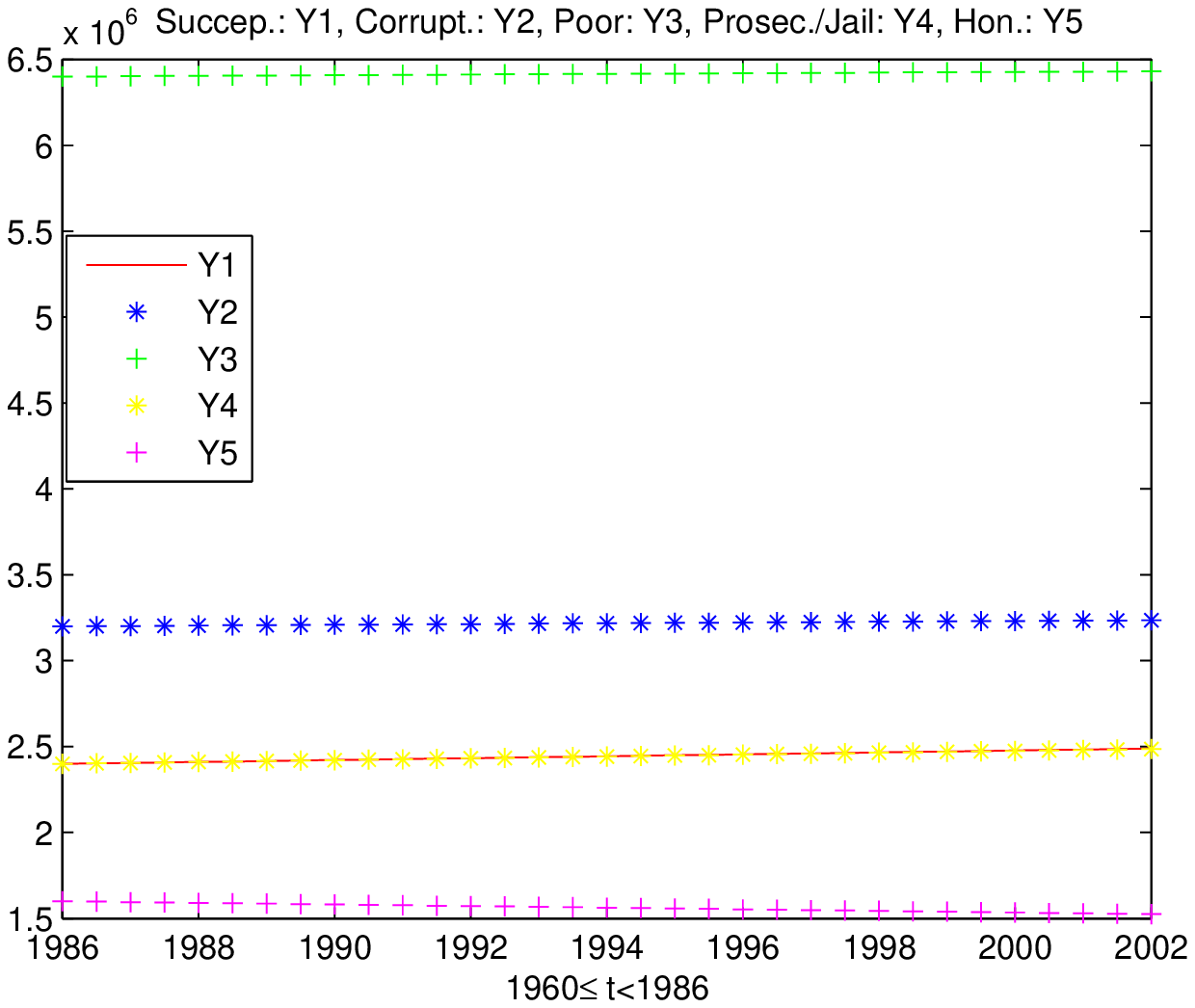,width=12cm}\\
         \end{tabular}
        \end{center}
        \caption{Analytical solution (y: in green), Computed solution (Y: in blue), Error (E: in red) for Example 4}
          \label{fig4}
          \end{figure}

\begin{figure}
         \begin{center}
         Situation of corruption and poverty in Cameroon: $2002$-$2022$.
          \begin{tabular}{c }
         \psfig{file=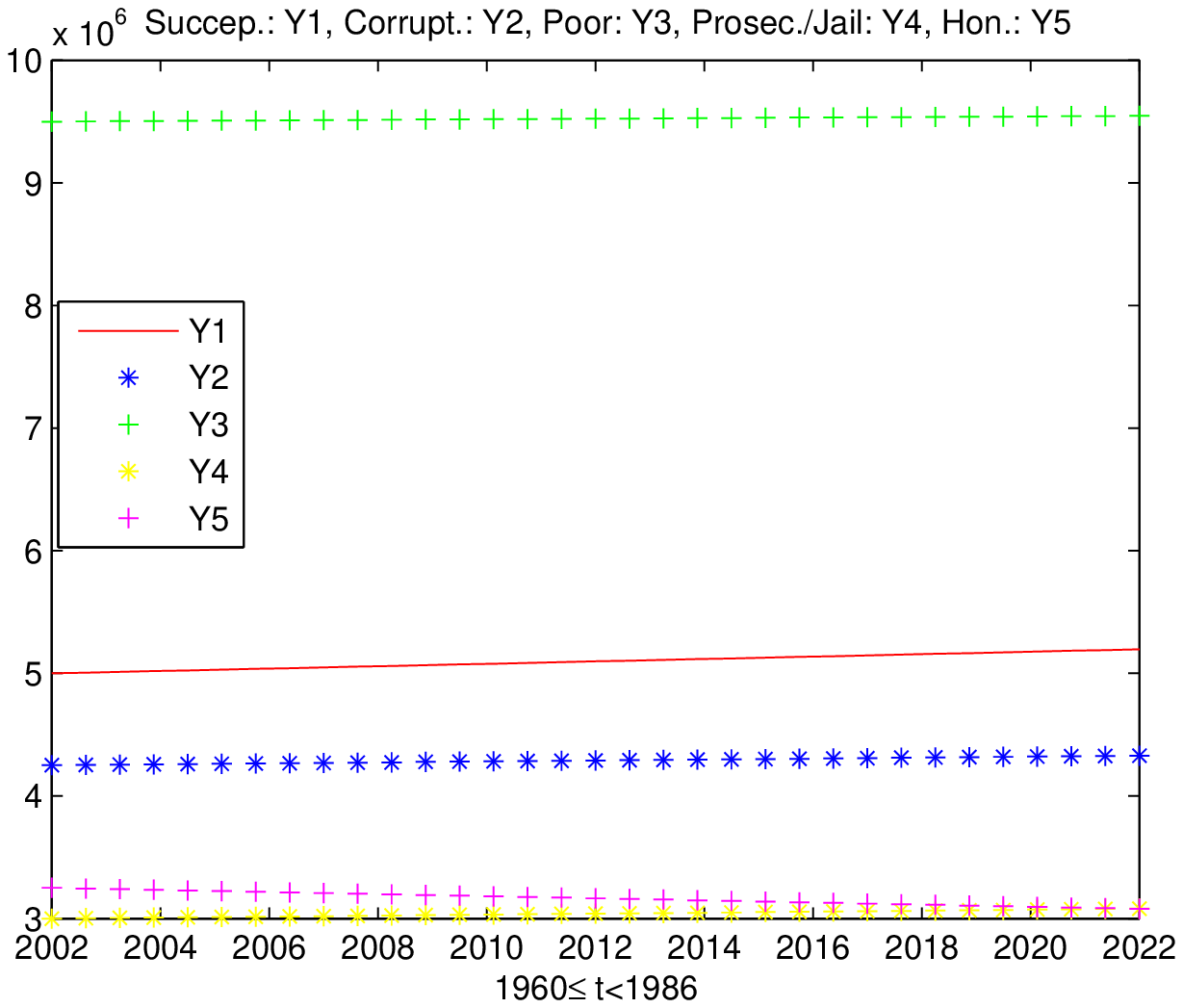,width=15cm} \\
         \end{tabular}
        \end{center}
        \caption{Exact solution (y: in green), Numerical solution (Y: in blue), Error (E: in red) for Example 5}
          \label{fig5}
          \end{figure}

         In this paper, we have developed a third-step second-order explicit numerical approach in a computed solution of a mathematical model on the dynamic of corruption and poverty $(\ref{1})$-$(\ref{5})$, subjects to initial condition $(\ref{6})$. The particular case on the dynamic of corruption and poverty in Cameroon has been discussed in the Section $\ref{sec5}$ (Numerical experiments). The theory has suggested that the proposed numerical technique is second-order convergent and stable for any value of the initial datum (Theorem $\ref{t}$). This result is confirmed by some numerical examples (see, Tables $1$-$2$ and Figures $\ref{fig1}$-$\ref{fig2}$). Furthermore, the study indicates that the new algorithm is fast and more efficient than a broad range of numerical schemes largely studied in the literature for solving such systems of nonlinear differential equations \cite{en1,en2,encovid,enarxivcovid,4nrn,16ca,10ca}. Finally, the developed numerical method is used to investigating and predicting the dynamic of corruption and poverty in Cameroon. The results provided by Tables $3$-$5$ and Figures $\ref{fig3}$-$\ref{fig5}$ confirm those discussed in \cite{gnm,8gnm,38gnm,35gnm,36gnm,37gnm}. Our future works will consider the development of a third-step third-order explicit method to assessing the dynamic of the poverty with effect of alcohol taking into account the particular case of Cameroon.\\

         \textbf{Acknowledgment.}  This work has been partially supported by the Deanship of Scientific Research of Imam Mohammad Ibn Saud Islamic University (IMSIU) under the Grant No. 331203.

            \newpage

     \end{document}